Original Article

# Laboratory study of mineral dust optical properties extending in the far-infrared spectral range

Perla Alalam[a,b,*], Hervé Herbin[c,*], Denis Petitprez[d], Sylvie Gosselin[d], Valentine Bizet[d], Paola Formenti[e], Pasquale Sellitto[a,f], Claudia Di Biagio[e]

[a]Univ Paris Est Creteil and Université Paris Cité, CNRS, LISA, F-94010 Créteil, France ; [b]CNES (Centre National d'Etudes Spatiales), 75001, Paris, France; [c]Université de Lille, CNRS, UMR 8518, LOA, F-59000 Lille, France; [d]Université de Lille, CNRS, UMR 8522, PC2A, F-59000 Lille, France; [e]Université Paris Cité and Univ Paris Est Creteil, CNRS, LISA, F-75013 Paris, France;[f]Istituto Nazionale di Geofisica e Vulcanologia, , Osservartorio Etneo, Catania, Italy

*CONTACT Perla Alalam perla.alalam@univ-lille.fr; Hervé Herbin herve.herbin@univ-lille.fr

**ABSTRACT**

Mineral dust is one of the most abundant atmospheric aerosol types and remains a major source of uncertainty in estimates of aerosol radiative forcing. This uncertainty is particularly important in the far-infrared (FIR) spectral region, where the optical properties of natural mineral dust, especially its complex refractive index (CRI), remain poorly known. Previous measurements have commonly relied on pellet- or film-based methods, which can alter particle size, shape, and vibrational modes, inducing errors in the CRI determination. In this study, we present a laboratory investigation on six natural mineral dust samples with contrasting compositions and originated from major dust source regions worldwide, including Arizona, Atacama, China, Niger, Saudi Arabia, and Tunisia. Suspended dust aerosols were generated under controlled dry conditions and characterized using extinction spectroscopy from the infrared 250 to 4000 cm−1 (2.5 to 40 µm) to the Ultraviolet-Visible (UV-Vis) from 10000 to 40000 cm−1 (0.25 to 1 µm) spectral range, combined with particle size distribution measurements. The real and imaginary parts of the CRI were retrieved for all six samples, revealing a significant absorption and pronounced spectral variability in the FIR. These variations should be accounted for in remote sensing and dust radiative effect studies.

## 1. Introduction

Mineral dust, lofted into the atmosphere by wind erosion over arid and semi-arid regions, constitutes one of the most abundant components of atmospheric aerosols and plays a major role in the Earth's radiative balance (Miller et al. 2014; Kok et al. 2026). Despite decades of research, quantifying aerosol radiative forcing remains one of the largest sources of uncertainty in climate studies (Forster et al. 2021). One important contributor to this uncertainty is the limited knowledge of aerosol spectral optical properties, particularly the complex refractive index (CRI). The CRI provides the fundamental link between the physicochemical properties of aerosol particles and their interaction with radiation. Accurate CRI data are essential for simulating dust–radiation interactions and for retrieving dust properties from remote-sensing observations (Sokolik and Toon 1999; Klüser et al. 2015; Alalam et al. 2022).

The spectral optical properties of mineral dust vary because its physicochemical characteristics differ among source regions and evolve during atmospheric transport (Di Biagio et al. 2023). The spectral response of mineral dust is governed by both its microphysical properties, including particle size distribution and morphology, and its chemical properties, particularly mineralogical composition and mixing state (Castellanos et al. 2024). The size distribution of emitted dust particles reflects the characteristics of source soils as well as the mechanics of wind-driven emission processes (Mahowald et al. 2014). It typically extends from submicron particles to several tens of micrometres in diameter (e.g., Ryder et al., 2013). In parallel, mineralogical composition varies significantly across source regions, with silicates, carbonates, and iron oxides contributing in different proportions at regional and global scales (Shen et al. 2009; Di Biagio et al. 2017).

Spaceborne infrared observations provide a powerful means of monitoring the dust life cycle from emission to transport and deposition at regional and global scales. However, the retrieval of dust optical depth, particle size distribution, altitude, and mineralogical information depends critically on accurate and representative aerosol optical properties, particularly the CRI (Sokolik et al. 1998). High spectral resolution instruments such as the Infrared Atmospheric Sounding Interferometer (IASI) have demonstrated the capability to retrieve aerosol properties in the mid-infrared domain (MIR) (Ginoux et al. 2012; Clarisse et al. 2019; Alalam et al. 2024).

Although satellite studies have demonstrated the potential of the MIR for dust detection and characterization, aerosol–radiation interactions at longer thermal-infrared wavelengths remain considerably less explored, particularly in the far-infrared (FIR, approximately 15–100 μm). This knowledge gap is important because a significant fraction of the Earth's outgoing longwave radiation is emitted in the FIR (Harries et al. 2008). Mineral dust absorption in the FIR can be comparable in magnitude to that observed in the MIR, indicating that FIR dust spectral signatures are not negligible (Di Biagio et al. 2026). The FIR may also contain spectral information that is less accessible in the MIR, because some minerals have no features in the MIR but strong absorption at longer wavelengths. This is particularly relevant for iron-oxides such as hematite, whose spectral signatures are only detectable in the FIR (Zhang et al. 2015). Detecting and characterizing dust in this spectral region therefore requires accurate knowledge of FIR optical properties.

This spectral region will become increasingly important with the advent of the forthcoming satellite missions the Infrared Atmospheric Sounding Interferometer - New Generation (IASI-NG) and Far-infrared Outgoing Radiation Understanding and Monitoring (FORUM), that will provide unprecedented hyperspectral observations across the MIR and FIR domains at high spectral resolution. IASI-NG is designed to improve spectral resolution and radiometric performance in the MIR down to ~ 650 $cm^{-1}$ (~ 15 μm), enhancing sensitivity to aerosol spectral signatures (Crevoisier et al. 2014), while FORUM will deliver the first global spectrally resolved measurements of outgoing longwave radiation down to ~100 $cm^{-1}$ (~100 μm) (Ridolfi et al. 2020). Recent sensitivity studies using pseudo-observations have highlighted the potential of FORUM to detect and quantify dust aerosols in the far infrared, showing that radiance sensitivity to dust optical depth in the FIR can be similar to that in the MIR and emphasizing the importance of accurate dust optical properties across the full FIR (Sellitto et al. 2026). Fully exploiting these missions therefore requires aerosol optical properties that are representative of atmospheric particle states across the entire thermal infrared spectrum.

Despite the radiative importance of the FIR and the increasing availability of spectrally resolved observations, current knowledge of the FIR optical properties of mineral dust remains limited. Historically, most available information has been derived from laboratory measurements of individual mineral components. As recently synthesized by Di Biagio et al. 2026, spectral datasets for major dust-forming minerals, including silicates, carbonates, and iron oxides, have been obtained from transmission or reflectance measurements of pellets, powders, and single crystals, in some cases extending down to wavenumbers of ~ 100 $cm^{-1}$ (~ 100 μm) (e.g., Spitzer and Kleinman, 1961; Querry, 1987; Glotch et al., 2007; Chehab et al., 2024). However, these studies generally focused on pure mineral phases rather than compositionally complex natural dust mixtures. As a result, the FIR spectral behavior of realistic atmospheric dust remains comparatively poorly constrained. The few available pellet-based studies of natural dust reported continuous absorption below 400 $cm^{-1}$ (beyond 25 μm), but only a few studies extended down to ~ 250 $cm^{-1}$ (~ 40 μm) (Fouquart et al. 1987; Volz 1972, 1973). More recently, Di Biagio et al., (2026) characterized natural dust samples and showed that absorption in the 400–650 $cm^{-1}$ range (15–25 μm) is non-negligible and strongly dependent on mineralogical composition. Nevertheless, pellet preparation may alter both the physical state and optical response of dust particles because the particles are embedded and compressed within a matrix rather than measured as freely suspended aerosols (Tamanai et al. 2006).

In this work we provide laboratory spectroscopic characterization of six mineral dust samples with contrasting mineralogical compositions measured directly in their suspended aerosol state. The aerosols are generated under controlled laboratory conditions and characterized in terms of their extinction spectra from 250 to 4000 $cm^{-1}$ (2.5 to 40 μm) and their particle size distributions. The measurements are then used to retrieve the spectral complex refractive indices of these samples, with particular emphasis on the FIR region down to 250 $cm^{-1}$. The paper is organized as follows: **Section 2** describes the experimental methodology, including the aerosol generation procedure, the extinction and size-distribution measurements, and the inversion framework used to retrieve the CRI. **Section 3** presents the main results of the study, from the measured extinction spectra and size distributions to the retrieved complex refractive indices and their uncertainty analysis, with particular emphasis on the new information obtained in the FIR. Finally, **Sect. 4** synthesizes the main conclusions.

## 2. Experimental methods

The experimental study was conducted using the spectroscopy platform of the PhysicoChimie des Processus de Combustion et de l'Atmosphère (PC2A) and Laboratoire d'Optique Atmosphérique (LOA). The experimental setup, previously detailed by Chehab et al., 2024, includes upgraded features from Herbin et al., 2023 and Hubert et al., 2017. Here, only the components and procedures relevant to the present study are summarized.

As shown in **Figure 1**, the setup consists of three main parts: the aerosol generation system (black dashed box), the optical properties measurements section (red dashed box), and the size distribution measurements section (violet dashed box). The different sections are connected using silicone tubes from TSI Inc. (inner diameter 4.8 mm, outer diameter 9.5 mm), with a total cumulative tubing length of approximately 1.6 m. The tubing is arranged as straight as possible to minimize bends and thereby reduce particle losses during aerosol transport through the experimental system.

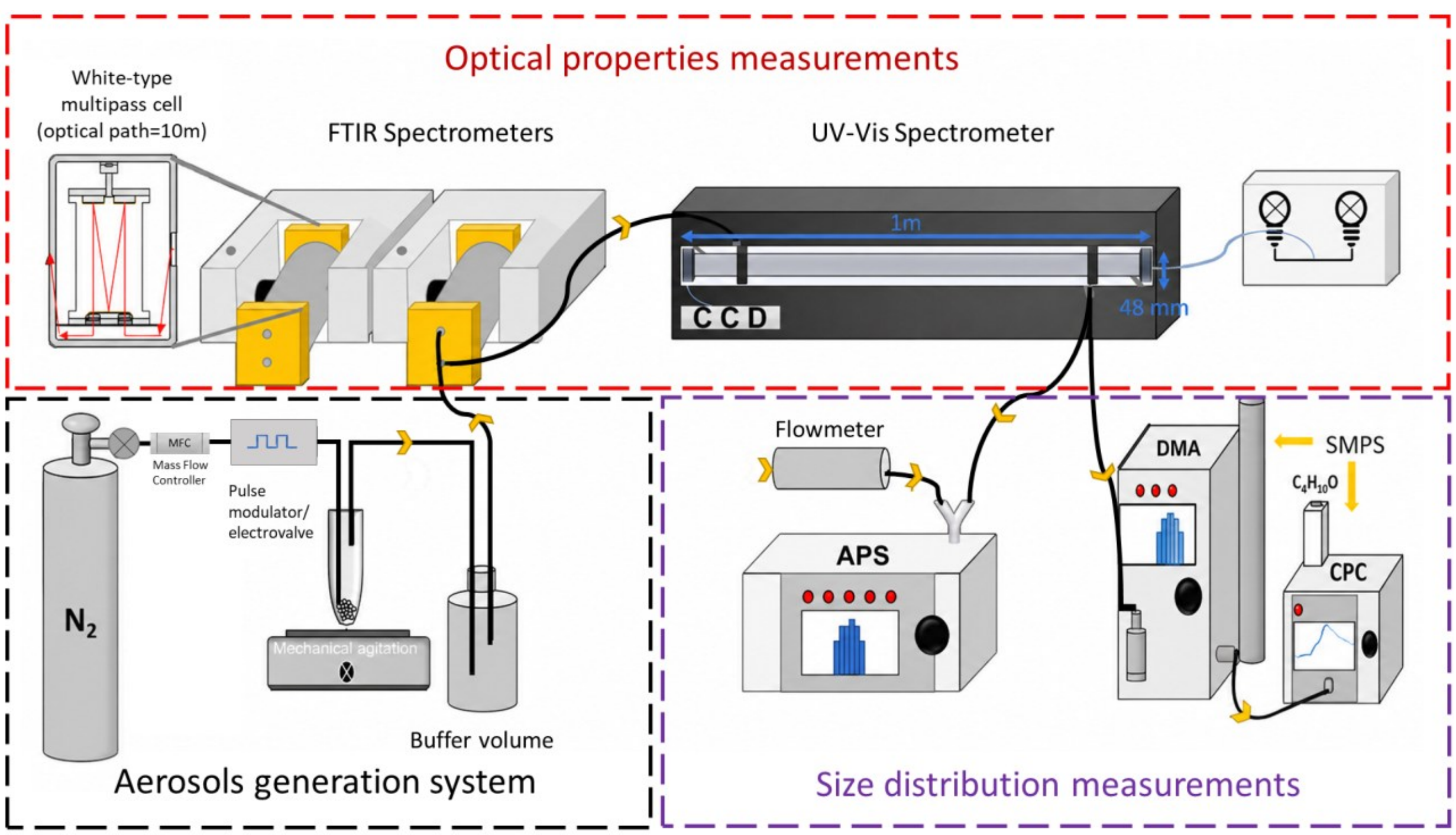

### *2.1. Aerosol generation system*

Samples used in the experiments were sieved to <1000 μm to remove non-erodible grains and heated at 353 K for 24 h to eliminate residual water, without altering their mineral structure (Sertsu and Sánchez 1978; Di Biagio et al. 2017). For each experiment, 4 g of soil was placed in a glass reactor. A magnetic stirring system operated at 1250 revolutions per minute (rpm) was used to generate dust aerosols. In addition, short pulses of nitrogen ($N_2$, purity 99.996 %) were introduced through an electronically controlled valve installed on the nitrogen inlet line upstream of the glass reactor. The pulsed flow enhanced soil-aerosol generation without requiring an increase in the initial sample mass, which was particularly advantageous for samples available only in limited quantities. Compared with mechanical agitation alone, it also improved the stability of the aerosol flow and particle number concentration during spectral acquisition. Pulse frequencies ranging from 1 to 15 $s^{-1}$ were selected for each sample according to its erodibility, in order to achieve comparable aerosol particle number of concentration among experiments. Across measurements, the aerosol flow was directed through a of 1 L buffer volume positioned downstream of the reactor and upstream of the measurement instruments. The buffer was used to homogenize and stabilize the aerosol flow and to prevent pressure fluctuations within the spectrometers. The aerosol concentration remained stable within 5% variability over the duration of the optical and particle-size measurements.

### *2.2. Optical measurements: Extinction Spectra*

 The aerosol extinction spectra were measured using a Thermo Fisher FTIR Nicolet iS50 spectrometer equipped with a 10 m White-type multipass cell. The system was equipped with interchangeable beam splitters, detectors, and cell windows to cover a broad spectral range. FIR spectra ($100-650\mathrm{cm}^{-1}, 15-100\mu\mathrm{m}$) were recorded at a resolution of 2 $\mathrm{cm}^{-1}$ using a deuterated triglycine sulfate (DTGS) detector with a solid-substrate beam splitter and polyethylene windows. MIR spectra ($650–4000 cm^{-1}, 2.5–15 \mu m$) were measured at a resolution of 0.5 $cm^{-1}$ using a mercury cadmium telluride (MCT) detector with a KBr beam splitter and $BaF_2$ windows.

After passing through the FTIR system, the aerosol entered a single-pass cell (1 m long and 48 mm in diameter) situated within a UV–Vis spectrometer. This system was equipped with two light sources: a halogen lamp covering $6667–27778\ cm^{-1}$ ($360 – 1500$ nm) and a deuterium lamp covering $25000–47619\ cm^{-1}$ ($210 – 400$ nm). Extinction measurements were recorded at a spectral resolution of 1 nm. **Table S.2.** in Supplementary materials summarizes the configurations of the detectors, beam splitters, and windows used in the spectrometers.

A single spectral acquisition is completed within 3 min, with the acquisition duration primarily constrained by the particle sizer measurement time. This duration is optimized to maximize spectrometers sensitivity and achieve a high signal-to-noise ratio SNR ($\geq 100$), while limiting particle deposition on the mirrors and cells windows.

The spectral extinction coefficient $\beta_{ext}(\nu)$ was calculated from the measured absorbance $A(\nu)$ and the optical path length $L$, as follows:

$$\beta_{ext}(\nu) = \frac{ln(10)A(\nu)}{L}, \quad (1)$$

The spectral measurements were performed under ambient laboratory conditions, with temperatures ranging from 290.7 to 293.9 K and atmospheric pressures from 0.963 to 1.009 atm. The same measurement protocol was followed for each experiment. Prior to the optical measurements, the spectrometers and optical cells were purged with $N_2$ to reduce absorption by water vapor, carbon dioxide, and other gaseous species along the optical path. For this purging procedure, $N_2$ was supplied at a flow rate of 3 L $min^{-1}$ using a mass flow controller (MFC). The progress of the purging procedure was monitored through the evolution of the background signal recorded by the spectrometers. Spectral acquisition was initiated once the background signal had stabilized. The purging procedure typically lasted 4h for the MIR window and up to 6h for the FIR. This procedure established dry conditions within the purged optical system, with a relative humidity (RH) of $\leq 20\%$.

An instrumental background spectrum was first recorded to estimate the baseline signal and instrumental noise. An aerosol-free pre-blank was then recorded under the same experimental conditions as the sample measurement. Aerosol generation was then initiated by activating the mechanical agitation and the pulse valve. Once a stable particle concentration was reached, the extinction spectrum and the corresponding particle size distribution were then measured simultaneously. Each spectral acquisition lasted 3 min, corresponding to the averaging interval applied to the particle size distribution measurements. Following the aerosol measurement, a second aerosol-free blank, hereafter referred to as the post-sample blank, was recorded to identify any residual extinction resulting from particle deposition on the optical cell mirrors. It was also used to assess the stability of the optical baseline and verify that the alignment of the FTIR components remained unchanged during the experiment. The complete measurement procedure was repeated several times for each dust sample to ensure measurement reproducibility.

Although the purge process significantly reduced gas-phase absorption bands, residual spectral lines may still persist. This may partly result from the ability of mineral aerosol particles to desorb water vapor from their surfaces (Chen et al. 2020) . Residual narrow gas-phase absorption features were reduced using the Savitzky-Golay filtering procedure, which applies local polynomial regression to each data point, generating a smoothed less distorted spectrum (Savitzky and Golay 1964). The smoothing effectively diminishes the amplitude of high-frequency signals arising from gas absorption lines within aerosol extinction spectra and was previously used with high spectral resolution data treatment (Ruffin et al. 2008). Therefore, this method was employed to finely distinguish aerosol absorption peaks caused by minerals molecular assignments and minimize interference from gas bands, while preserving both the shape and intensity of the aerosol bands.

To calculate the uncertainties associated with the spectral measurements, FTIR instrumental noise was derived from the manufacturer's photometric specifications and scaled to the acquisition conditions, yielding an absorbance uncertainty of $\pm(2-4)\times 10^{-4}$. This corresponded to an extinction uncertainty of $(1.5-3)\times 10^{-5}\mathrm{m}^{-1}$ for fixed 10 m path length defined by the manufacturer. As this value cannot be adjusted by the user, its contribution to the measurement uncertainty was considered negligible.

In the UV–Vis, the median absorbance noise was $1\times 10^{-4}$, corresponding to an extinction uncertainty of $2\times 10^{-4}\ \mathrm{m}^{-1}$ and a propagated relative uncertainty of approximately 0.5%. Additional uncertainties associated with gas-lines removal and spectral smoothing were estimated from the difference between raw and smoothed spectra, yielding median smoothing-induced uncertainties of $6\times 10^{-5}\ \mathrm{m}^{-1}$ in the FIR, $1\times 10^{-5}\ \mathrm{m}^{-1}$ in the MIR, and $8\times 10^{-6}\ \mathrm{m}^{-1}$ in the UV–Vis. Overall, median relative uncertainties remained below 0.5% over the measured regions, while the extinction coefficients $\beta_{ext}\,(\nu)$ were estimated to remain within approximately 1% over the full spectral range.

### *2.3. Size distribution measurements*

The aerosol number distribution $n(D)$ was measured using two complementary instruments, that operate in different size ranges based on different measurement principles. Fine-particle size distributions were measured using a Scanning Mobility Particle Sizer (SMPS, TSI, combining a DMA Model 3080 and a CPC Model 3775 with a 3.0/0.3 $\mathrm{L\cdot min^{-1}}$ sheath flow/aerosol flow rate) at a time resolution of 3 min. The SMPS measures the electrical mobility diameter $D_m$ of particles in the nominal range of 14 nm to 600 nm, defined as the diameter of a spherical particle exhibiting the same migration velocity in a uniform electric field as the particle being analysed. Multiple charge correction (MCC) and diffusion corrections were applied to the SMPS size distributions using the internal algorithms of the TSI Aerosol Instrument Manager (AIM) software. Coarse particles were measured using an Aerodynamic Particle Sizer (APS, TSI Model 3321 with a 5 $\mathrm{L\cdot min^{-1}}$ of aerosol flow rate) that operates based on time-of-flight measurements of individual particles at time resolution of 1 min. The APS determines the aerodynamic diameters $D_a$ in the nominal range of 0.523 to 20 µm. The aerodynamic diameter $D_a$ is defined as the equivalent diameter of a spherical particle with unit density $\rho_p$ settling through air at the same terminal velocity as the measured particle. Note that three consecutive APS distributions were averaged to obtain a size distribution corresponding to each 3 min of SMPS acquisition.

Similarly to the optical measurements, particle size distributions were recorded during the pre- and post-sample blank measurements to evaluate the presence of residual airborne particles in the experimental system. Once consistency among the replicate measurements had been confirmed, one representative dataset including an aerosol extinction spectrum and the corresponding particle-size distribution acquired over the same 3 min interval, was selected for further analysis. The selection was based on the overall quality and stability of both measurements, including the absence of instrumental artefacts, abnormal baseline variations, and unstable particle concentrations during the acquisition period.

*2.3.1. Particle shape factor correction*

Since the SMPS and APS measure mobility diameter $D_m$ and aerodynamic diameter $D_a$, respectively, both are converted to geometric diameter $D_g$, defined here as the volume-equivalent diameter. The conversions equations accounted by the particle density $\rho_p$ and the dynamic shape factor $\chi$ are described by DeCarlo et al., (2004), and are given by:

$$D_g = D_a\sqrt{\frac{\chi\rho_0}{\rho_p}}, \qquad (2)$$

$$D_g = \frac{D_m}{\chi}, \qquad (3)$$

where $\rho_0$ is the density of air of 1 $\mathrm{g.cm^{-1}}$ , $\rho_p$ is the effective density of pure minerals densities calculated based on values reported from the literature (See **Table S.3** and **Table S.4**). The dynamic shape factor was assumed $\chi = 1.19$, corresponding to an average value previously estimated for mineral dust (Huang et al. 2021; Formenti and Di Biagio 2024). **Equations (2)** and **(3)** assume that the Cunningham slip correction factor $C_s$ for both the aerodynamic and mobility diameters is equal to unity, following the approximation proposed by Kim et al., (2005).

The uncertainty associated with the conversion from instrument-reported diameter to $D_g$ was evaluated separately for the SMPS and APS. For the SMPS, propagation of the uncertainty in the dynamic shape factor resulted in a relative geometric-diameter uncertainty of approximately 12% in $D_g$. For the APS, propagation of the uncertainties in both particle density and dynamic shape factor resulted in a relative geometric-diameter uncertainty of approximately 6% in $D_g$.

Uncertainties in $dN/d\log D_g$ were evaluated independently from the diameter uncertainties and for each size bin because it depends on the number of particles detected within that bin. For the SMPS, the uncertainty in $dN/d\log D_g$ included contributions from particle-counting statistics, CPC aerosol-flow uncertainty, CPC counting efficiency, and the diffusion correction. For the APS, the uncertainty in $dN/d\log D_g$ included contributions from counting statistics and aerosol-flow uncertainty. **Table S.7** summarizes the uncertainty budget associated with the SMPS and APS measurements.

*2.3.2. Particle losses correction*

Because the optical and size-distribution measurements are performed at different positions located sequentially along the same aerosol flow, particle transport losses can modify the aerosol size distribution between the generation system, the optical cells, and the SMPS–APS sampling point. Correcting for these losses is therefore necessary to estimate the particle size distribution representative of the aerosol interacting with the optical beam and to reduce biases in the retrieved complex refractive indices.

To obtain an experimental assessment of particle losses along the setup, the SMPS–APS sampling system was connected successively at different locations within the experimental apparatus, and the corresponding particle size distributions were compared. However, because the measurements at the different positions could not be performed under strictly identical aerosol-generation and concentration conditions, this approach provides primarily a qualitative evaluation of the size dependence of the particle losses. It was therefore used as an independent consistency check of the size dependence of the losses predicted by the numerical calculations, rather than to derive the quantitative correction applied to the measured size distributions.

Quantitative particle-losses calculations were performed using the Particle Loss Calculator software (PLC; Von der Weiden et al., 2009), which estimates aerosol transport losses resulting from the main deposition mechanisms occurring along the sampling line. The input parameters include the geometry of the experimental system tubes (e.g. inner and outer diameters, tube inclination and bends), the sampling orientation and flow rate, the particle density $\rho_p$ and the dynamic shape factor $\chi$.

The calculated particle losses exhibited a pronounced dependence on particle diameter, with enhanced losses for both ultrafine and coarse particles. Diffusional deposition resulted in losses exceeding 40% for $D_g \leq 0.05$ µm. The losses then decreased to a minimum of approximately 4% near $D_g = 0.2$ µm, before increasing in the coarse particle range because of gravitational settling and inertial deposition. Particle losses reached approximately 27% at $D_g \sim 1.0$ µm, 50% at $D_g \sim 1.4$ µm, 77% at $D_g \sim 1.9$ µm, and 93% at $D_g \sim 2.3$ µm (See **Figure S.2**). Since the SMPS and APS were located downstream of the FTIR cell, only particle losses occurring between the FTIR optical interaction region and the SMPS–APS sampling point were considered in the correction.

To retrieve the aerosol size distribution representative of the FTIR optical interaction region, the combined SMPS–APS size distribution $dN/dlogD_g$, was corrected for particle losses using the following equation:

$$\left[dN/dlogD_g\right]_{Corr} = \left(dN/dlogD_g\right)/\left(1 - \left(PLC\left(D_g\right)/100\right)\right) , \quad (4)$$

The uncertainty associated with the PLC correction of $\left[dN/dlogD_g\right]_{Corr}$ was evaluated through a sensitivity analysis in which the PLC input parameters were varied within their estimated uncertainties. The individual sensitivity contributions were combined quadratically for each particle-size bin to obtain the total uncertainty in the calculated loss function. The resulting lower and upper loss estimates were then applied to the measured size distribution, resulting in asymmetric lower and upper uncertainties in the corrected size distribution $\left[dN/dlogD_g\right]_{Corr}$.

The relative uncertainty in the corrected size distribution attributable to the PLC correction was generally below 3% for Dg<0.5 µm, with values below 1% across much of the 0.03–0.30 µm. It increased to approximately 14% at $D_g \sim 1$ µm, 37% at $D_g \sim 1.4$ µm, and 95% at $D_g \sim 2.0$ µm. Size bins for which the nominal PLC-calculated losses were 95% were excluded from the subsequent analysis because the correction becomes highly sensitive to small uncertainties as the loss fraction approaches unity.

*2.3.3. Multimodal lognormal fit*

Following Seinfeld and Pandis, (2016), the particle size distribution $n(D)$ is described mathematically by the sum of lognormal distributions, and it is given by:

$$n(D) = dN/dlogD = \sum_{i=1}^{m} \frac{N_i}{log(\sigma_{g,i})\sqrt{2\pi}} exp\left(-\frac{\left(log(D) - log(D_{g,i})\right)^2}{2\,log(\sigma_{g,i})^2}\right), \quad (5)$$

where $m$ is the total number of modes in the size distribution, $D$ represents the equivalent volume diameter or the geometric diameter of particles. $N_i$ , $D_{g,i}$ and $\sigma_{g,i}$ are the number concentration, the geometric mean diameter, the geometric standard deviation of particle diameters in the $i^{th}$ mode, respectively.

The corrected size distributions were fitted to a multimodal log-normal model from **equation (5)** using nonlinear least squares regression. Each log-normal mode $i$ was parameterized by its initial guess for $(N_i, D_{g,i}, \sigma_{g,i})$ with bounds defined to constrain physically reasonable values for each mode. Four modes were fitted based on the shape of the size distributions: an ultrafine mode (UFM), a fine mode (FM), a coarse mode 1 (CM1) and a coarse mode 2 (CM2). Fitted results from individual modes were summed to reconstruct the total combined distribution and model performance was quantified using the coefficient of determination ($R^2$) and the root mean square error (RMSE). Parameter uncertainties were estimated from the covariance matrix of the nonlinear regression and ranged between 20 and 30%. These uncertainties were used to define the a priori covariance matrix of the modal parameters for the optical adjustment of the size distribution described in **Sect. 2.4.1**.

***2.4. Complex refractive index retrieval***

Starting with the measured extinction spectra and aerosol size distribution, the real and imaginary parts of the CRI denoted as $m(\nu) = n(\nu) + i\kappa(\nu)$, are retrieved over the measured spectral range. Retrieving the CRI from extinction measurements constitutes a mathematically ill-posed inverse problem because extinction at each wavenumber $\nu$ depends simultaneously on both $n$ and $\kappa$. An a priori estimate is therefore required before iterative adjustment toward the optimal solution. In the inversion process, the measurement vector **y** corresponds to the experimental extinction $\beta_{ext}(\nu)$, and the state vector **x** corresponds to the CRI unknowns $n(v)$ and $\kappa(v)$. The complete workflow used to process the optical measurements and incorporate them into the complex refractive index retrieval procedure is illustrated in **Fig 2**.

*2.4.1. Size distribution adjustment*

Despite the corrections described above, the particle size distribution used in the CRI retrieval remains affected by uncertainties associated with the conversion to geometric diameter, particularly through the assumed dynamic shape factor and particle density and the particle-loss correction calculated with the PLC (Reid et al. 2003; Huang et al. 2021). Therefore, the consistency of the corrected SMPS–APS size distribution was additionally evaluated using the measured extinction spectrum.

The modal parameters obtained from the multimodal fit described in **Sect. 2.3.3** were used as the a priori size-distribution parameters, together with their associated uncertainties. The size-distribution parameters were then adjusted within their uncertainty ranges by minimizing the difference between the extinction spectrum calculated using Mie theory and the measured extinction spectrum. The optimization was performed between 1600 and 40000 $\mathrm{cm}^{-1}$, while spectral regions dominated by strong mineral vibrational absorption bands were excluded. The ultrafine-mode parameters were kept fixed during the adjustment because this mode has a negligible contribution to the extinction over the spectral interval considered, whereas the parameters of the fine and coarse modes were allowed to vary within their a priori uncertainty ranges. The mathematical formulation of the adjustment procedure is provided in **Appendix A**.

The resulting distribution is not interpreted as an independent or intrinsically more accurate measurement of the particle size distribution. Rather, it represents an optically constrained effective size distribution that maintains consistency between the particle-size measurements and the observed extinction spectrum. This optical consistency is required to retrieve a spectrally coherent complex refractive index over the investigated spectral range, which is particularly important for radiative-transfer calculations and remote-sensing applications.

*2.4.2. Numerical model*

Once the adjusted size distribution and its errors are estimated, it is used as input for the inversion step. The inversion is initiated using the Rayleigh approximation over the infrared spectral region where absorption dominates, so that the extinction coefficient can be approximated by the absorption coefficient, $\beta_{ext}(v) = \beta_{abs}(v)$. In practice, this range was limited to the region below the extinction minimum located around $1600\ \mathrm{cm}^{-1}$, after the last strong MIR absorption band.

In addition, the aerosol concentration inside the multi-pass White cell was sufficiently low to maintain optically diluted conditions. Consequently, particle–particle interactions and multiple scattering were considered negligible, and the measured attenuation was assumed to result from independent single-particle extinction events within a homogeneous, well-dispersed aerosol population. Under these assumptions, following Dartois and Bauerecker, (2008) and Herbin et al., (2017), for the particle's dispersion the imaginary index $\kappa(v)$ represents the extinction of a dispersed homogeneous medium, and it is given by:

$$A_i(\nu) = \frac{\beta_{ext}(v)}{6\pi\, v\, f_v L}, \tag{6}$$

where $f_v = (\pi/6) \sum_i C_i D_i^3$ is the volume fraction of the particles.

Having obtained the imaginary part $A_i(\nu)$, the real part $A_r(\nu)$ can be estimated using the causality principle following the Kramers-Kronig relations. However, limitations of these formulas must be considered due to integrals over an infinite wavenumber range. Bachrach and Brown, (1970) developed the Single Subtractive Kramers-Kronig (SSKK) relation to reduce errors caused by a spectrum with finite wavenumbers. It improves the accuracy of the Kramers-Kronig estimation by subtracting the integral at a known reference point, known as an "anchor point" where the real refractive index value is measured independently.

Applying the Kramers–Kronig integral over the finite spectral interval covered by the experiment may introduce boundary-related errors and lead to inconsistencies in the retrieved complex refractive index. To reduce these edge effects, the imaginary component $\kappa(\nu)$ is extrapolated beyond the measured spectral limits, following approaches previously used for aerosol optical constant retrievals (Tisdale et al. 1998; Biermann et al. 2000).

The extrapolation depends on the optical behavior of the suspended aerosol and can be chosen to represent constant, increasing, or decreasing spectral behavior outside the measured interval (See Herbin et al., 2017)

After applying the extrapolation, the real component is calculated using the SSKK relation, and it is given by:

$$A_r(\nu) \approx A_0(\nu) + \frac{2(\nu^2 - \nu_0^2)}{\pi} \mathrm{P} \int_{\nu_l}^{\nu_u} \frac{\nu' A_i(\nu')}{(\nu'^2 - \nu^2)(\nu'^2 - \nu_0^2)} \mathrm{d}\nu, \tag{7}$$

where $A_0(\nu) = 1 - (3/(2 + \mathrm{n}(\nu_0)^2))$, $\mathrm{n}(\nu_0)$ is the real index value at the anchor point corresponding to the wavenumber $\nu_0$, $\nu_l$ and $\nu_u$ are the lower and upper wavenumber limits of the experimental spectral range. In this study, the anchor point $\mathrm{n}(\nu_0)$ is fixed at $\nu_0$ =10000 cm$^{-1}$ using literature values from Di Biagio et al., (2019).

The complex dielectric function is be written as:

$$\epsilon(\nu) = \frac{1+2A(\nu)}{1-A(\nu)}, \tag{8}$$

$$\begin{cases} \epsilon_r(\nu) = \frac{3(1-A_r(\nu))}{(1-A_r(\nu))^2 + A_i(\nu)^2} - 2 \\ \epsilon_i(\nu) = \frac{3A_i(\nu)}{(1-A_r(\nu))^2 + A_i(\nu)^2} \end{cases}, \tag{9}$$

Using $\epsilon_r = n^2 + \kappa^2$ and $\epsilon_i = 2n\kappa$, the real and imaginary parts are then obtained as:

$$\begin{cases} n(\nu) = \left[\frac{(\epsilon_r(\nu)^2 + \epsilon_i(\nu)^2)^{1/2} + \epsilon_r(\nu)}{2}\right]^{1/2} \\ \kappa(\nu) = \left[\frac{(\epsilon_r(\nu)^2 + \epsilon_i(\nu)^2)^{1/2} - \epsilon_r(\nu)}{2}\right]^{1/2}, \end{cases} \tag{10}$$

The retrieved $\mathrm{n}(\nu)$ and $\kappa(\nu)$ are used to simulate an extinction spectrum $\beta_{ext,sim}(\nu)$ and calculate the difference with the experimental spectrum $\beta_{ext,exp}(\nu)$, given by:

$$\Delta A_i(\nu) = \frac{\beta_{ext,exp}(\nu) - \beta_{ext,sim}(\nu)}{6\pi\, \nu\, f_v L}, \tag{11}$$

The values of $A_i(\nu)_j$ are then adjusted iteratively by minimizing the difference between the simulation and the experimental spectra by adding $\Delta A_i(\nu)_j$ to the next iteration $A_i(\nu)_{j+1}$, calculated by:

$$A_i(\nu)_{j+1} = A_i(\nu)_j + \Delta A_i(\nu)_j, \tag{12}$$

To provide a spectrally continuous starting point for the OEM retrieval, the spectral values of $n(\nu)$ and $\kappa(\nu)$ were first iteratively adjusted by minimizing the difference between simulated and measured extinction spectra. At each iteration, the $A_i(\nu)$ was updated according to **Eq. (12),** from which updated values of $n(\nu)$ and $\kappa(\nu)$ and a new extinction spectrum were obtained. Convergence was evaluated using the root mean square error ($RMSE = \sqrt{\Sigma\left(y_i - F(x_i)\right)^2/n}$) computed over $n$ =3900 spectral channels. The $n(\nu)$ and $\kappa(\nu)$ values corresponding to the minimum $RMSE$ were then retained as the reference state for the OEM retrieval, which employs Mie theory as the forward model. In this inversion step, this reference state is treated as a regularization state $\mathbf{x}_{\mathrm{reg}}$ rather than a strictly measurement-independent Bayesian a priori. This distinction is made because $\mathbf{x}_{\mathrm{reg}}$ is obtained through the preliminary Rayleigh-based iterative convergence and therefore dependent on the experimental measurement. In addition, the non-retrieved parameter vector **b** encompasses the multimodal size-distribution parameters obtained from **Sect.2.4.1**. Therefore, the preliminary step provides a spectrally continuous reference state for the CRI OEM retrieval, since an optical anchor point is used in the preliminary SSKK calculation, it is not imposed as a fixed constraint during the OEM retrieval. Finally, the total CRI uncertainty $\sigma_{\mathrm{x,CRI}}$ is calculated from the propagated contributions of the regularization constraint, measurement noise, and uncertainties in the non-retrieved size-distribution parameters. Comprehensive mathematical formulations, covariance definitions, and uncertainty propagation details are all provided in the Supplementary Material.

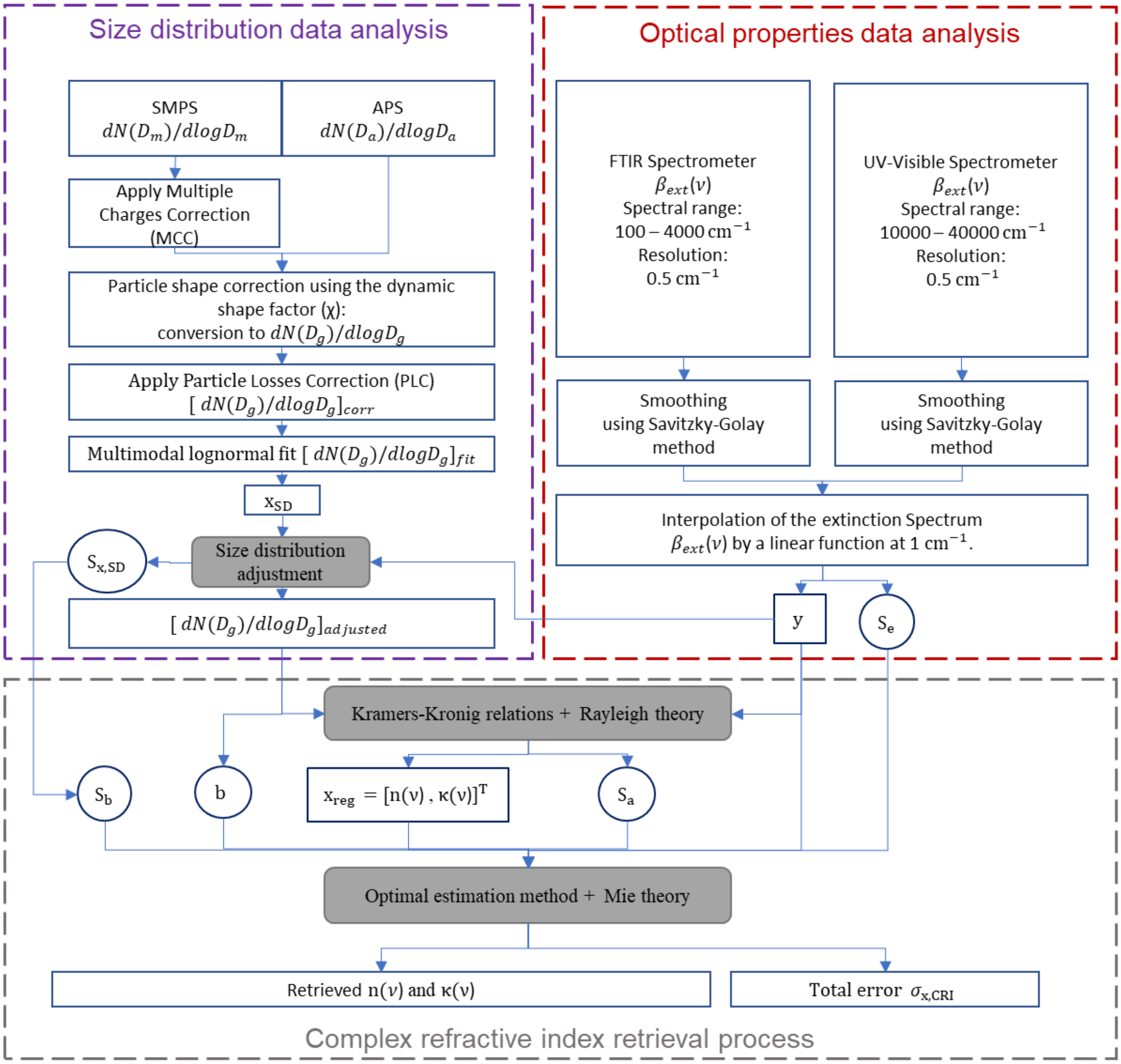

## 3. Results

### *3.1. Number size distribution*

**Figure 3** shows the corrected aerosol number size distributions, $[dN(D_g)/dlogD_g]_{corr}$ (black dots) with the total uncertainties (gray error bars), obtained for the six source samples with the multimodal combined fit, $[dN(D_g)/dlogD_g]_{fit}$ (magenta solid line) and the optically adjusted distribution $[dN(D_g)/dlogD_g]_{adjusted}$ (orange solid line). The detailed retrieved modal parameters, associated posterior uncertainties, and relative shifts with respect to the a priori state are provided in the **Supplement Materials**.

The a priori multimodal fits reproduce the corrected size distributions reasonably well with $R^2$ values ranging from 0.80 to 0.98. All samples exhibit a multimodal structure composed of an ultrafine mode, a fine mode, and two coarse modes. In all cases, the fine mode constitutes the dominant contribution to particle number concentration $N$ with geometric diameters $D_g$ centered between 0.27 and 0.37 μm and peak concentrations ranging from approximately $4.7 \times 10^4$ to $1.9 \times 10^5$ particles $cm^{-3}$. The ultrafine mode, observed below 0.03 μm, contributes less strongly to the total particle number concentration but remains detectable in all distributions, with relatively enhanced intensity for Tunisia and Saudi Arabia. The coarse fraction is generally represented by two broader modes, with the first one centered approximately on 0.75 μm and exhibiting relatively similar peak concentrations across most samples (about $1.0 \times 10^4$–$1.5 \times 10^4$ particles $cm^{-3}$), except for Niger and Saudi Arabia, where lower concentrations close to $4.7 \times 10^3$ particles $cm^{-3}$ are observed. The second coarse mode occurs at larger diameters, with geometric diameters ranging from approximately 4 to 10 μm, and remains much weaker in number concentration than the first coarse mode, with peak values varying from about 1 to 73 particles $cm^{-3}$. Its contribution is most pronounced for Arizona and China, whereas it remains weaker for the other samples.

The optically adjusted distributions reproduce the overall measured multimodal structure of the corrected SMPS–APS distributions. The adjustment modifies both the characteristic sizes and the relative number concentrations of the different modes. Fine-mode concentrations increase systematically in most samples, while coarse-mode concentrations show larger relative variability but remain associated with low absolute concentrations. This does not imply larger uncertainty in the fine-mode measurements, but rather reflects the redistribution of particle number required by the optical inversion. The inversion preserves an overall multimodal structure while modifying the relative contribution of the different modes. The results are consistent with differences in the coarse aerosol population being partially compensated by an increase in retrieved fine mode. Therefore, the adjusted distribution should be interpreted as an optically constrained representation of the aerosol size distribution.

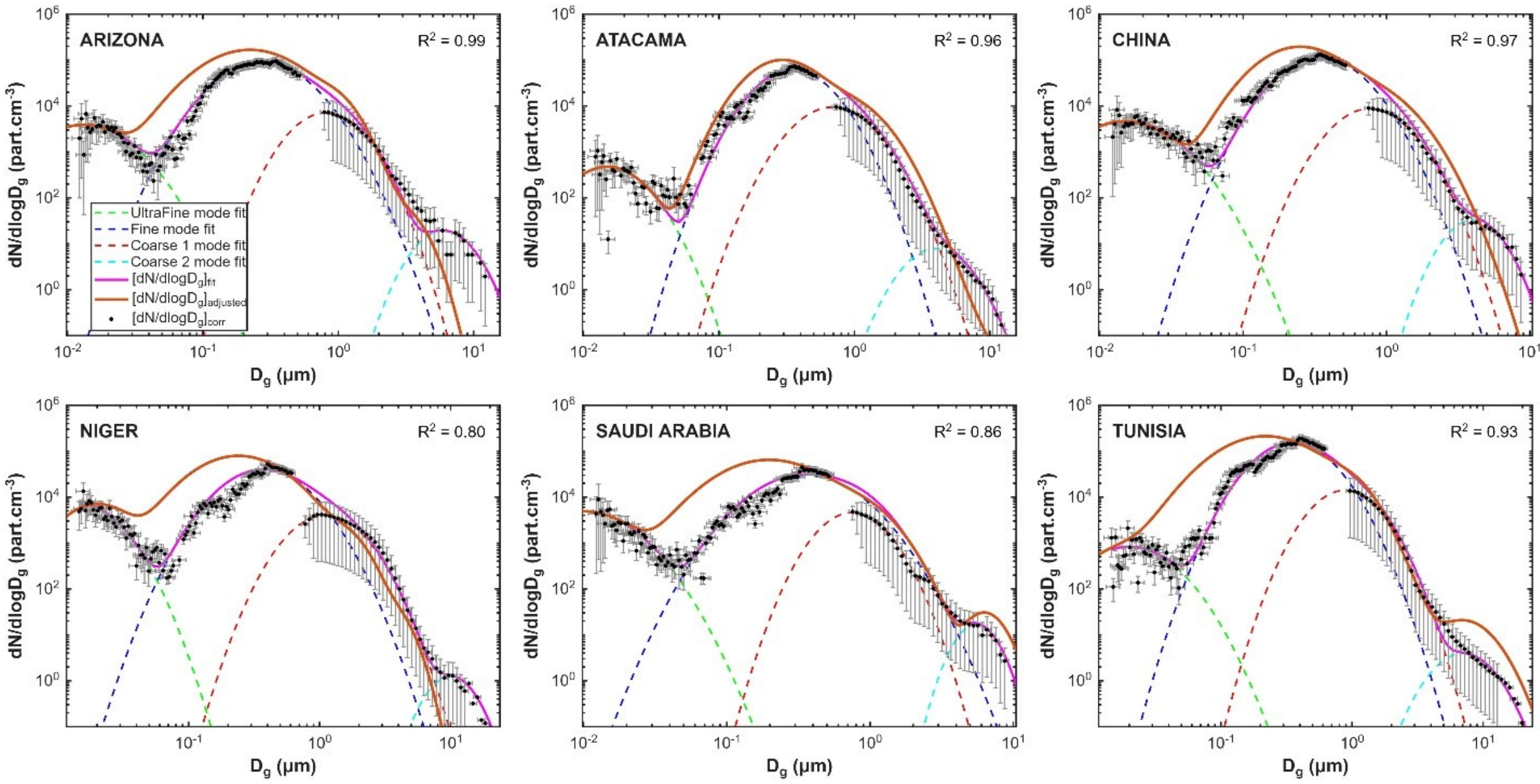

### *3.2. Aerosols extinction spectra*

The smoothed extinction spectra of the six experimentally generated dust samples (Arizona, Atacama, China, Niger, Saudi Arabia, and Tunisia) are illustrated in **Figure 4**. The application of smoothing reduces gases perturbations in the raw spectra and allows clearer identification of mineral-specific absorption features while preserving the principal spectral signatures. Nevertheless, measurements below 250 $cm^{-1}$ (> 40 μm) remain limited by residual water vapour interference after smoothing and by consequence, these are not exploitable for CRI retrievals. The observed differences in extinction intensities mainly result from the specific number concentration and size distribution of the generated aerosols. Despite these intensity variations, all spectra display a common overall structure extending from FIR to the UV, with local differences in spectral band expression reflecting contrast in mineralogical composition. Below approximately 1600 $cm^{-1}$ (above 6.25 μm wavelength), extinction is dominated by absorption features associated with mineral vibrational modes characteristic of the various dust minerals. Above 1600 $cm^{-1}$ (below 6.25 μm), the spectra progressively lose distinct band structure and evolve toward smoother extinction behavior, indicating increasing dominance of scattering. Although, some persistent absorption features near 3800 $cm^{-1}$ are observed, indicating residual O–H stretching contributions. In the shortwave infrared (SWIR) region (between 4000 and 10000 cm⁻¹), no direct experimental measurements were performed resulting in a discontinuity between the measured spectral domains.

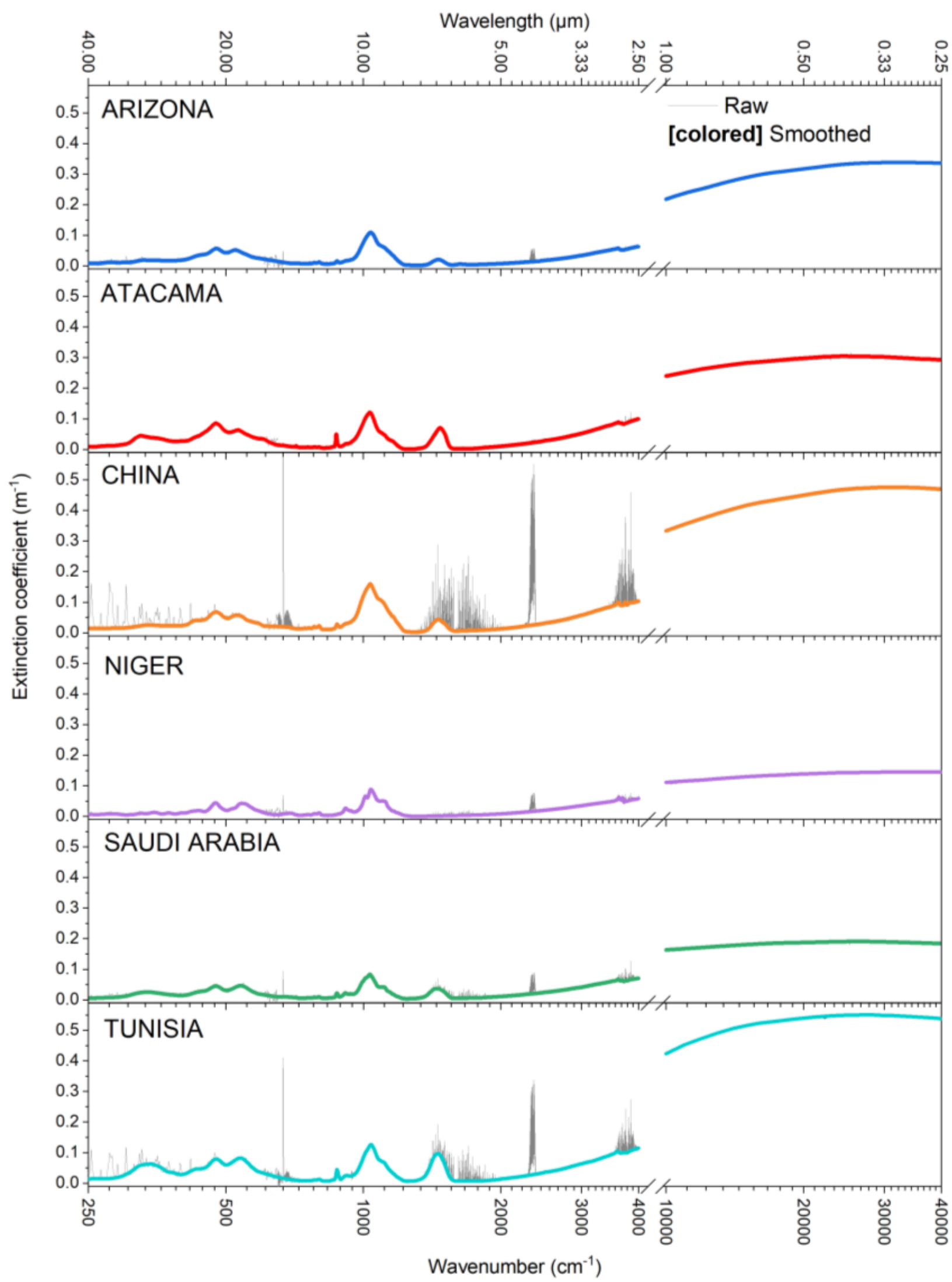

**Figure 5** illustrates the main molecular absorption features observed in the infrared range between 250 and 1600 cm⁻¹. In the 650–1600 cm⁻¹, the spectra show the characteristic absorption bands already reported for these samples by Di Biagio et al., 2017, with composition-dependent differences mainly controlled by silicate and carbonate minerals. The dominant feature, observed between approximately 1000 and 1100 cm⁻¹, and corresponds to Si–O / Si–O–Si stretching vibrations associated with silicate minerals mainly clays and quartz (Balan et al. 2001; Müller et al. 2014; Madejová et al. 2017). Carbonates exhibit characteristic absorption features near 875 cm⁻¹, assigned to C–O bending, and between 1400 and 1500 cm⁻¹, corresponding to asymmetric C–O stretching, both typical of calcite-rich compositions (Campbell and Poduska 2020). These bands are particularly enhanced in Tunisia and Atacama, whereas Niger shows almost no carbonate-related absorption in this region, consistently with its mineralogical composition (Di Biagio et al., 2017). Minor absorption near 915 cm⁻¹ is attributed to -OH bending modes associated with clay minerals (Madejová et al. 2017).

The extension of the measurements in this work below 666 cm⁻¹ allows the spectral signature of suspended dust to be explored in a lower-wavenumber infrared region that has been less documented for airborne mineral dust. In this range, additional absorption structures are observed, including bands between 450 and 550 cm⁻¹, associated with Si–O–Si and Si–O–Al bending modes, particularly in clay-rich samples (Madejová et al. 2017; Farmer 1974). Lower-wavenumber features below 400 cm⁻¹ include contributions from Ca–O lattice vibrations (Campbell and Poduska 2020). Hematite-related Fe–O vibrations are expected between approximately 250 and 560 cm⁻¹, with characteristic bands reported in this region (Onari et al. 1977; Glotch et al. 2006). However, these iron-oxide features are not clearly resolved in the present spectra, most likely because iron oxides occur only in minor proportions across all samples and their absorption bands are masked by stronger silicate and carbonate lattice vibrations in the same spectral region. The characteristic wavenumber positions and vibrational assignments of the principal minerals commonly present in desert dust are summarized in **Table S.8**.

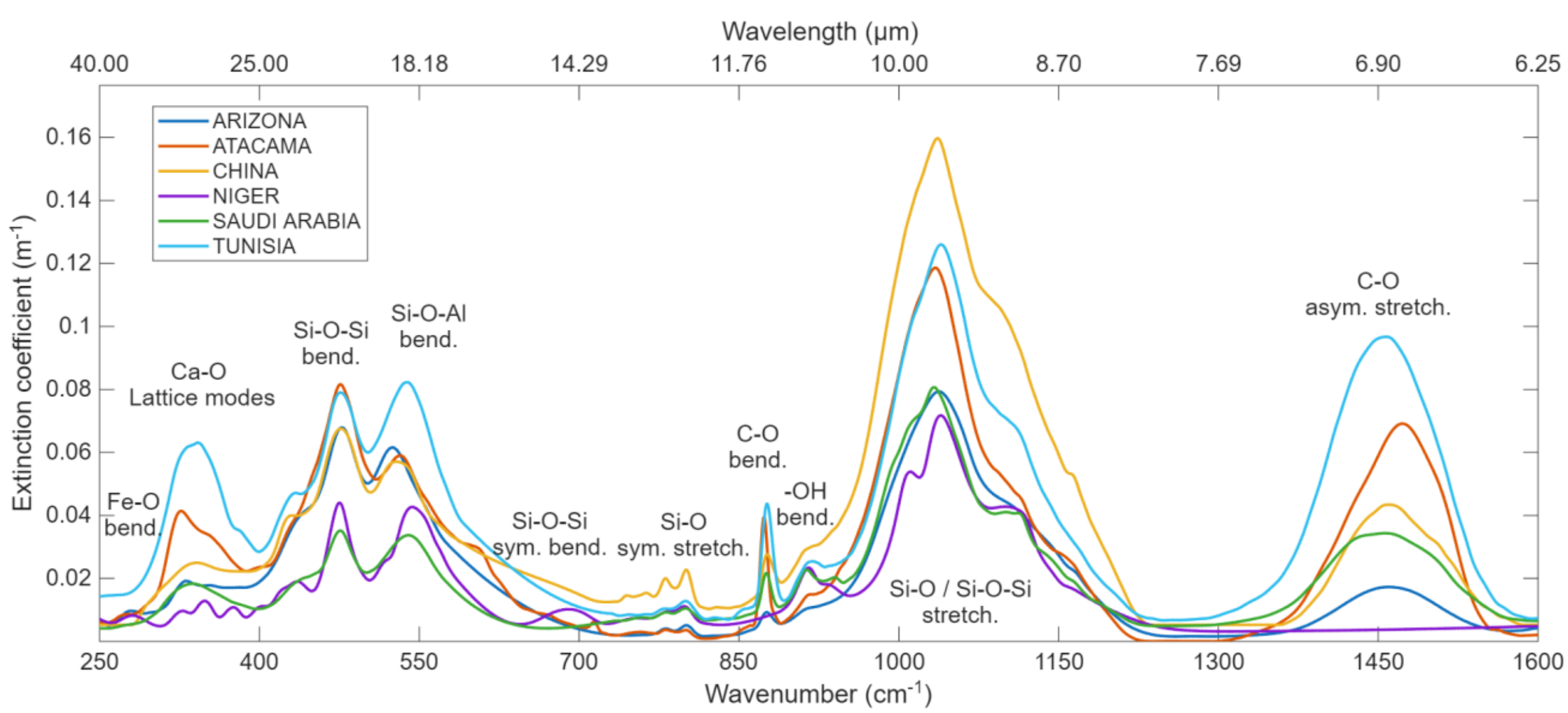

### *3.3. Aerosols complex refractive indices and discussion*

**Figure 6** presents the retrieved CRI for the six dust samples, showing the spectral evolution of the real part n(ν) and the imaginary part κ(ν) over the measured spectral range (250- 4000 $cm^{-1}$ and 10000-40000 $cm^{-1}$). As illustrated in the figure the CRI values retrieved in the FIR are systematically of the same order of magnitude as, and in some spectral regions larger than, those retrieved in the MIR. The imaginary part κ(ν) ranges from approximately 0.003 to 2.85 in the FIR below 650 $cm^{-1}$, whereas in the MIR it varies from values close to zero to approximately 1.64, with the largest values occurring in the vicinity of the main silicate absorption band around 1000–1050 $cm^{-1}$. In addition to its magnitude, the FIR optical response exhibits a rich spectral structure that differs considerably among the six samples. These source-dependent features provide complementary information to the MIR bands and may offer additional constraints on the mineralogical composition of atmospheric dust. In parallel, the real part n(ν) displays in both FIR and MIR the expected dispersive response associated with variations of the imaginary component, with local extrema typically varying between approximately 0.71 and 3.29 depending on the sample, before gradually converging toward nearly constant values between approximately 1.48 and 1.54 in the near-infrared, visible and UV spectral ranges.

The shaded envelopes shown in **Fig. 6** for the CRI represent the retrieval uncertainty estimated from the total covariance matrix $\mathbf{S_{tot}}$. Over most of the spectral domain, the relative uncertainty remains below 10 % for n(ν) and κ(ν), indicating stable convergence of the inversion procedure and strong consistency between measured and simulated extinction spectra. Larger relative uncertainties occur locally near sharp spectral structures, where extinction becomes more sensitive to simultaneous perturbations in CRI and particle-size parameters. According to the Rodgers, (2000) error decomposition, this behavior is mainly controlled by the forward-model contribution associated with uncertainties in the adjusted multimodal size distribution, whereas the measurement contribution remains comparatively small (about 1 %), due to the high signal-to-noise ratio of the extinction spectra, which is the main input information used in the algorithm. The a priori covariance matrix mainly acts as a regularization constraint and does not dominate the final posterior uncertainty. Importantly, the size-related contribution requires a specific interpretation. The experimentally measured particle size distribution provides the initial constraint for the retrieval, while the parameters of its multimodal representation are then adjusted within the inversion in order to optimize the agreement between measured and simulated extinction spectra. Therefore, the uncertainty associated with the adjusted size distribution does not represent the experimental measurement uncertainty, but rather the uncertainty of the size-distribution representation after spectral adjustment. A sensitivity example comparing the CRI uncertainty obtained by propagating the uncertainties of the measured and adjusted size distributions is provided in **Fig. S.6**. Finally, the fact that the uncertainty envelopes remain smaller than the inter-sample variability confirms that the observed regional differences are physically valid and cannot be attributed to inversion artefacts.

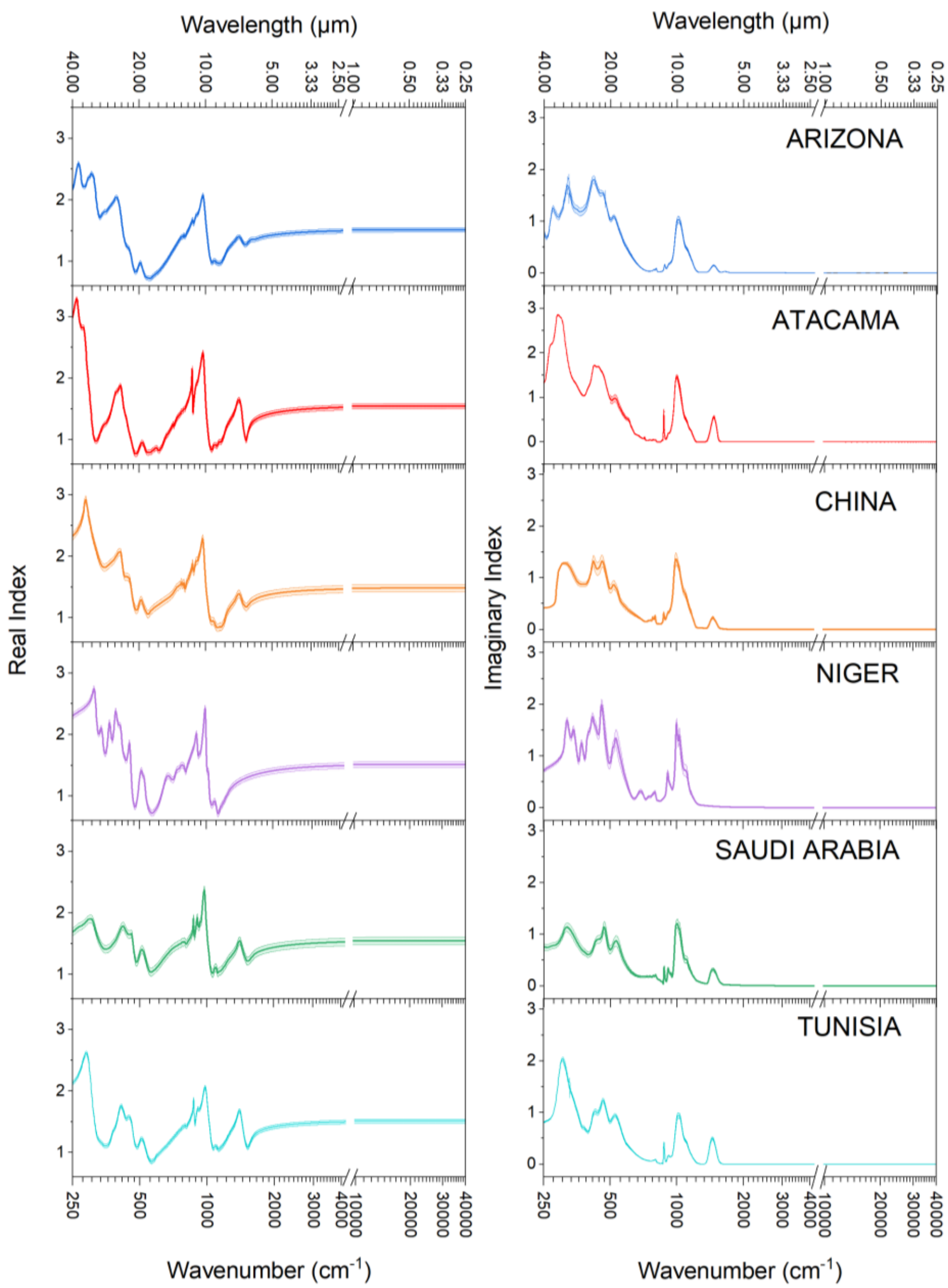

**Figure 7** provides comparison between the CRIs retrieved in the present study and values reported in the literature. The black curves represent the mean real and imaginary parts of the complex refractive index obtained for the six investigated dust samples, while the grey shaded areas indicate $\pm 1\sigma$ standard deviation around the mean and therefore illustrate the inter-sample variability. The comparison includes CRIs derived from pellet measurements of mineral-dust samples collected from different source regions or atmospheric environments, shown as dashed curves (Volz 1973; Fouquart et al. 1987; Hess et al. 1998; Di Biagio et al.

2014). The aerosol-phase measurements of Di Biagio et al., (2017), obtained from laboratory-resuspended dust originating from the same source materials investigated here, are shown as a solid purple curve.

This comparison is not intended as a direct quantitative evaluation because the datasets differ in several important respects, including measurement technique, sample composition, aerosol-size distribution, and spectral resolution. In particular, pellet measurements may be affected by particle compaction, interactions with the embedding matrix, and differences between bulk and aerosol-phase optical behaviour. Nevertheless, the comparison reveals several differences in spectral structure between the datasets. In particular, variations are observed around the main Si–O stretching band near 1000 $cm^{-1}$ and the carbonate-related C–O band around 1400–1450 $cm^{-1}$. Differences between the present results and those of Di Biagio et al. (2017), despite the use of dust from similar source materials, may reflect differences in the aerosol size distribution, which are accounted for in the retrieval algorithm. The size distributions generated in the present study contain a larger contribution from fine-particles compared with Di Biagio et al. (2017) having coarser distribution (see **Figure S.8** in Supplementary). In addition, as mineralogical components may be distributed differently among particle-size fractions, these differences can modify their optically weighted contribution as illustrated in **Figure S.9** and therefore affect the retrieved CRI. The results presented here therefore highlight the critical importance of the particle size distribution when retrieving the CRI, with particular attention to the accurate representation of the coarse fraction.

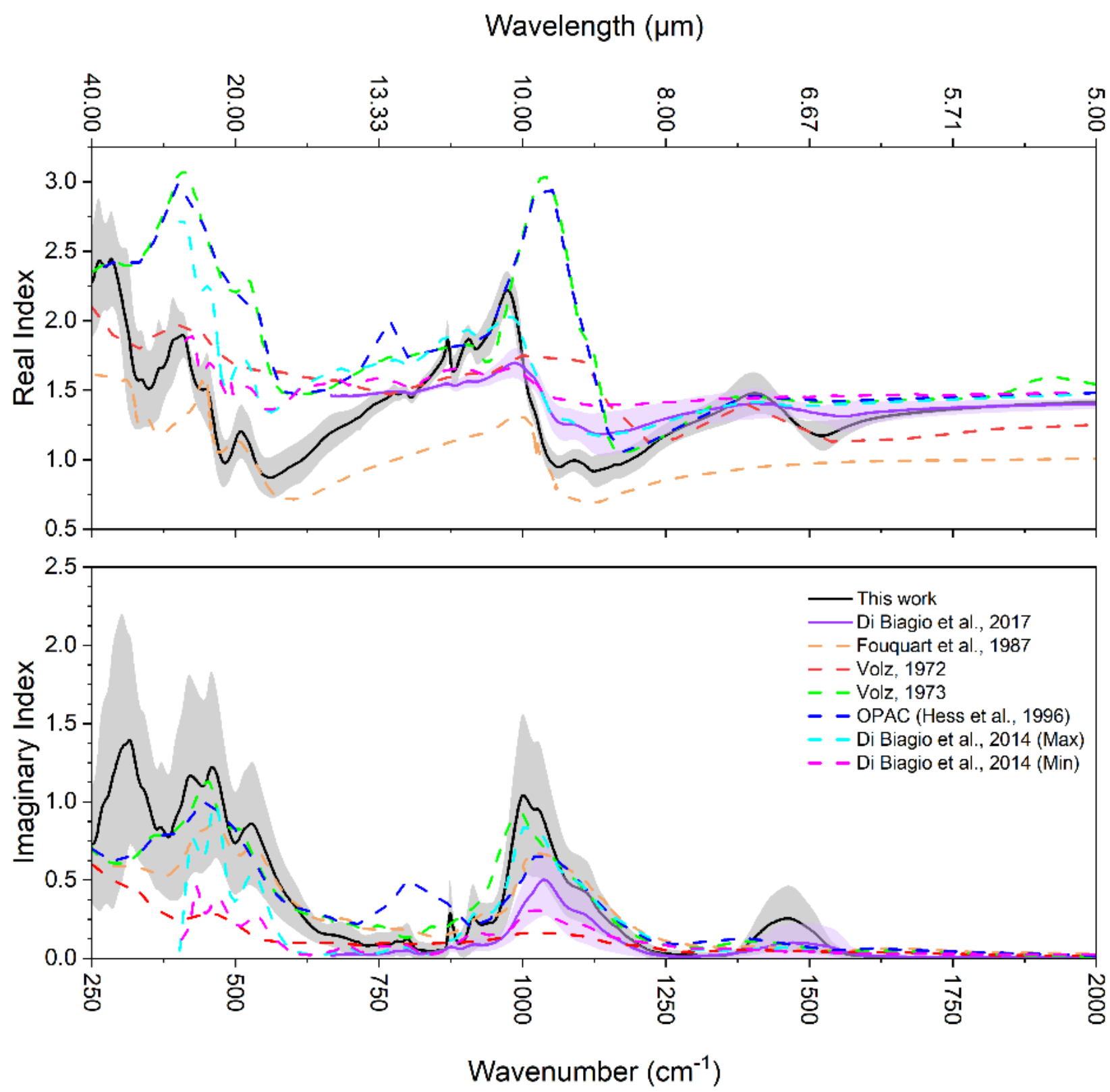

## 4. Conclusion

In this study, we performed laboratory measurements of extinction spectra and retrieved the CRIs of mineral dust aerosols from the FIR to the UV, using atmospherically suspended particles rather than particles embedded in pellets. The approach combines controlled aerosol generation, IR and UV–Vis extinction measurements, particle size distribution characterization, and an inversion framework based on subtractive Kramers–Kronig relationships and Mie theory.

The six investigated dust samples display clear source-dependent differences in both their extinction spectra and retrieved CRIs, with distinct spectral signatures observed throughout the FIR and MIR regions. These results confirm the relevance of dust spectral features beyond the MIR, as previously suggested by Di Biagio et al., (2026) from pellet measurements extending down to 400 cm$^{-1}$ (25 μm). By extending the measurements and CRI retrievals to 250 cm$^{-1}$ (40 μm) and considering suspended aerosol particles, we show that dust–radiation interactions remain significant well into the FIR. This spectral domain therefore provides complementary optical information, with structures and amplitudes that are sensitive to dust source and mineralogical composition. The results from this study, providing CRIs data across the MIR and FIR spectral domains represent a step forward in advancing our understanding of the spectral interaction of mineral dust with radiation and to support active and passive remote sensing. The availability of a spectrally extended set of CRIs support the joint interpretation of observations acquired by complementary techniques across different spectral ranges. Indeed, as demonstrated in Sellitto et al., (2026), the radiance sensitivity to dust optical depth in the FIR is comparable to that in the MIR, therefore supporting the potential to detect and quantify dust aerosols based on the forthcoming FORUM observations and its synergy with IASI-NG. Future work should extend this dataset to a broader range of dust sources and cover the entire FIR range (up to 100 µm), as well as to assess the radiative effects across the whole mid and far infrared range under representative atmospheric conditions

## 5. Appendix: Optimal estimation method

The Optimal Estimation Method (OEM) framework described by Rodgers, 2000 was used for both the optical adjustment of the particle size distribution and the retrieval of the CRI. The two inversions use the same general formalism but different state and non-retrieved parameter vectors.

For measurement vector $\mathbf{y}$, the forward model is written as:

$$\mathbf{y} = \mathbf{F}(\mathbf{x}, \mathbf{b}) + \boldsymbol{\epsilon}, \quad \text{(A.1)}$$

where $\mathbf{x}$ is the retrieved state vector, $\mathbf{b}$ contains the non-retrieved parameters of the forward model, and $\boldsymbol{\epsilon}$ is the measurement error.

Because the retrieved parameters cover several orders of magnitude, the inversion was performed in logarithmic state space. The solution $\mathbf{x}$ is found by the iterative process given by:

$$\ln(\mathbf{x_{i+1}}) = \ln(\mathbf{x_i}) + \left(\mathbf{K}^{\mathrm{T}}\mathbf{S_e}^{-1}\mathbf{K} + \mathbf{S_a}^{-1} + \gamma\mathbf{I}\right)^{-1}\left[\mathbf{K}^{\mathrm{T}}\mathbf{S_e}^{-1}\left(\mathbf{y} - \mathbf{F}(\mathbf{x_i})\right) - \mathbf{S_a}^{-1}\left(\ln(\mathbf{x_i}) - \ln(\mathbf{x_a})\right)\right], \quad \text{(A.2)}$$

where $\mathbf{K}$ is the Jacobian matrix of the forward model $\mathbf{F}(\mathbf{x_i})$ given by $\mathbf{K} = \frac{\Delta \mathbf{F}(\mathbf{x_i})}{\Delta \mathbf{ln x_i}}$ , $\gamma$ is the Levenberg–Marquardt damping parameter, and $\mathbf{S_e}$ and $\mathbf{S_a}$ are the measurement and a priori error covariance matrices, respectively.

For the particle size distribution adjustment, the damping term was used to stabilize convergence, whereas for the CRI retrieval the standard Gauss–Newton formulation was applied by setting $\gamma = 0$. The size distribution adjustment and CRI retrievals uncertainties are found in detail in the supplementary material.

## Data availability statement

The retrieved complex refractive index datasets and their associated uncertainties for the six mineral dust samples will be made publicly available through the AERIS data repository at the ACCRID database (https://accrid.loa.univ-lille.fr/) and AERIS (https://www.aeris-data.fr/icare/). The measured spectra and particle size distributions are available from the corresponding author upon request. The codes used for the retrieval and uncertainty analyses are laboratory specific routines under active development and are therefore not publicly archived.

## CRediT Roles

CDB, PS, and HH conceptualized and defined the main scientific objectives of the study. HH and DP designed the setup and developed the experimental methodology. PA performed the experiments with the technical support of VB and SG and contributions by HH, DP and CDB. PA carried out the data analysis, prepared the figures and wrote the original manuscript draft. CDB, HH, DP and PS provided guidance with the analysis and contributed to the interpretation of the study's results. PF and CDB provided the dust samples and mineralogical data. PA and HH contributed to the development and validation of the retrieval methodology. CDB and PS provided funding acquisition and project administration. All authors contributed to the discussion of the results and reviewed and edited the manuscript.

**Acknowledgments**

The authors strongly thank S. Alfaro, Meinrat O. Andreae, B. Chatenet, M. Kardous, R. Losno, B. Marticorena, J. L. Rajot, D. Seibert, and G. Vargas, who participated in the collection of the soil samples used in this study. The authors would like also to thank Adrien Deschamps who supported this collaboration through the CNES. In addition, authors would like to thank Quentin Libois and Laurent Labannote for their advices in the framework of the FORUM space mission.

**Funding**

The authors acknowledge financial support from the Centre national d'études spatiales (CNES), France (ROR: https://ror.org/04h1h0y33), within the framework of the FORUM space mission, the Programme National de Télédétection Spatiale (grant no. IR-DUST) and the European Space Agency (grant no. v4000151382/26/NL/IB/ar). P.A. was supported by the CNES through a postdoctoral fellowship. The French State under the France-2030 programme and the Initiative of Excellence of the University of Lille are acknowledged for the funding and support granted to the R-CDP-24-003-AREA project.

**Discloser statement**

The authors report there are no competing interests to declare.

**ORCID**

Will be added in list form by the typesetter if provided (not required).

**Captions**

Figure 1. Schematic representation of the experimental setup used in this work (adapted from Chehab et al. 2024). It is disposed of: an aerosol generation system (black dashed box), optical properties measurements with FTIR and UV-Visible spectrometers (red dashed box), and size distribution measurements (violet dashed box) using an Aerodynamic Particle Sizer (APS) and a Scanning Mobility Particle Sizer (SMPS) constituted of a Differential Mobility Analyzer (DMA) and a Condensation Particle Counter (CPC).

Figure 2. Schematic workflow for complex refractive index retrieval from size distribution and optical-property analyses. The workflow combines two analytical branches: size distribution data analysis (dashed violet), and optical properties data analysis (dashed red). These datasets are integrated through Kramers–Kronig relations with Rayleigh theory to obtain initial optical estimates, followed by iterative refinement using the optimal estimation method coupled with the Mie scattering model to retrieve the real $\boldsymbol{n}(\boldsymbol{\nu})$ and imaginary, $\boldsymbol{\kappa}(\boldsymbol{\nu})$ parts of the complex refractive index (dashed gray).

Figure 3. Number size distributions $\left[dN(D_g)/dlogD_g\right]_{corr}$ as function of the geometric diameter $D_g$ of aerosols dust samples from Arizona, Atacama, China, Niger, Saudi Arabia and Tunisia (black dots). The mobility diameter measured by the SMPS and the aerodynamic diameter measured by the APS were converted to geometrical diameter using the dynamic shape factor $\chi$= 1.19. Colored dashed lines represent the individual lognormal mode fits: ultrafine

(green dashed line), fine (blue dashed line), coarse 1 (red dashed line) and coarse 2 (cyan dashed line), the combined multimodal fit $\left[dN\left(D_g\right)/dlogD_g\right]_{fit}$ (magenta solid line) with the corresponding $R^2$ and the adjusted modes fit $\left[dN\left(D_g\right)/dlogD_g\right]_{adjusted}$ (red solid line) .

Figure 4. Smoothed extinction spectra of experimentally generated dust aerosols from Arizona, Atacama, China, Niger, Saudi Arabia, and Tunisia over the full measured spectral range. Gray curves represent raw extinction spectra, while colored curves show respectively the corresponding smoothed spectra after Savitzky–Golay filtering. Note that the discontinuity between 4000 and 10000 $cm^{-1}$ corresponds to the spectral interval without direct measurements.

Figure 5. Infrared extinction spectra of the six dust samples in the 250–1600 $cm^{-1}$ region highlighting the principal mineral absorption spectral features. Major absorption bands associated with silicates, carbonates, hydroxyl groups, and lattice vibrations are indicated.

Figure 6. Retrieved real n(ν) (left panels) and imaginary κ(ν) (right panels) parts of the complex refractive index for the six mineral dust samples as a function of wavenumber over 250–4000 $cm^{-1}$ and 10000–40000 $cm^{-1}$. Solid colored lines represent the retrieved CRIs, and shaded envelopes indicate the associated retrieval uncertainties, ranging from approximately 1 to 10 % for n(ν) and 1 to 15 % for κ(ν) depending on the spectral region and sample. The spectral discontinuity between 4000 and 10000 $cm^{-1}$ corresponds to the non-measured gap between the infrared and visible–near-infrared domains.

Figure 7. Spectral variation of the real, $\boldsymbol{n(v)}$ (upper panel), and imaginary, $\kappa(v)$ (lower panel), parts of the complex refractive index of mineral dust between 200 and 2000 $^{-1}$ (5–50 µm). The black curves represent the central values obtained in this work, while the grey shaded areas indicate the associated standard deviation of the datasets. The results are compared with the refractive-index datasets of Fouquart et al., 1987, Volz, 1972; 1973, Hess et al.**,** 1998 (known as OPAC), Di Biagio et al.**,** 2014 (minimum and maximum values), and Di Biagio et al. (2017).

# Supplementary Material

## Laboratory study of mineral dust optical properties extending in the far-infrared spectral range

Perla Alalam[a,b,*], Hervé Herbin[c,*], Denis Petitprez[d], Sylvie Gosselin[d], Valentine Bizet[d], Paola Formenti[e], Pasquale Sellitto[a,f], Claudia Di Biagio[e]

[a]Univ Paris Est Creteil and Université Paris Cité, CNRS, LISA, F-94010 Créteil, France ; [b]CNES (Centre National d'Etudes Spatiales), 75001, Paris, France; [c]Université de Lille, CNRS, UMR 8518, LOA, F-59000 Lille, France; [d]Université de Lille, CNRS, UMR 8522, PC2A, F-59000 Lille, France; [e]Université Paris Cité and Univ Paris Est Creteil, CNRS, LISA, F-75013 Paris, France;[f]Istituto Nazionale di Geofisica e Vulcanologia, , Osservartorio Etneo, Catania, Italy

# Supplementary A: Mineralogical composition information

**Figure S.1** presents the mineralogical distribution of the selected soils using a ternary representation based on three major mineral families: phyllosilicates (including clays), tectosilicates (including quartz and feldspars), and carbonates (including calcite and dolomite), which together account for the dominant mineral fractions in all samples. The ternary classification was derived from X-ray diffraction (XRD)-based mineral abundances, as presented in Di Biagio et al., (2017), while the detailed mineralogical composition and sample information are provided in **Fig. S1** and **Table S1** in the Supplement.

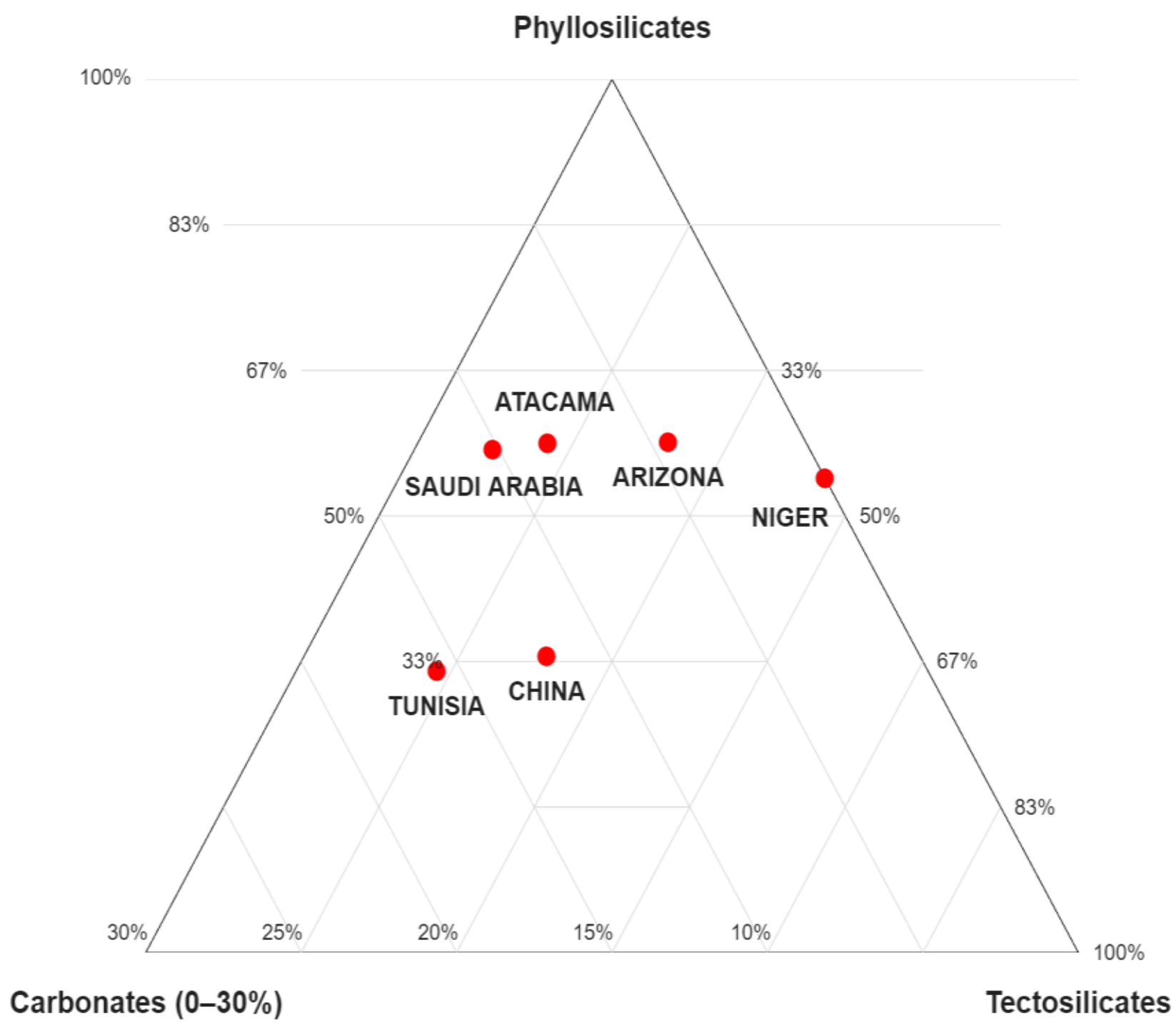

**Figure S.1:** Ternary diagram showing the mineralogical distribution of the six dust samples used in this study (Arizona, Atacama, China, Niger, Saudi Arabia, and Tunisia), based on X-ray diffraction mineral abundances reported by Di Biagio et al., 2017. The diagram groups the dominant mineral phases into three major mineral families: phyllosilicates (including clays), tectosilicates (including quartz and feldspars), and carbonates (including calcite and dolomite). The carbonate axis is limited to 30 % because carbonate fractions remain low in all samples. Minor iron oxide fractions (hematite and goethite), present only in small proportions, are not explicitly represented.

**Table S.1:** Information of soil samples used in this study (adapted from Di Biagio et al., 2017).

| Sample Name | Collection coordinates | Geographical zone | Country | Desert zone |
|---|---|---|---|---|
| ARIZONA | 33.15° N, 112.08° W | North America | United States of America | Sonoran Desert |
| ATACAMA | 23.70° S, 70.40° W | South America | Argentina | Patagonian Desert |
| CHINA | 39.43° N, 105.67° E | East Asia | China | Gobi Desert |
| NIGER | 13.87° N, 2.63° E | North Africa | Niger | Sahel (Banizoumbou) |
| SAUDI ARABIA | 27.82° N, 41.98° E | Middle East | Saudi Arabia | Nefud Desert |
| TUNISIA | 33.02° N, 10.67° E | North Africa | Tunisia | Sahara Desert (Maouna) |

# Supplementary B: Spectrometers configuration

**Table S.2:** List of each spectral range spectrometer's configuration: detector, beam splitter, windows and spectral resolution (adapted from Chehab et al., 2024).

| Spectral range ($cm^{-1}$) | Detector | Beam splitter | Windows of cell | Spectral resolution |
|---|---|---|---|---|
| 100–650 | DTGS | Solid substrate | Polyethylene | 2 $cm^{-1}$ |
| 650–4000 | MCT | KBr | $BaF_2$ | 0.5 $cm^{-1}$ |
| 9000–50,000 | CCD | | Quartz | 1 nm |

# Supplementary C: Particle losses analysis

To qualitatively assess particle losses along the experimental setup, the SMPS and APS were successively connected at three positions: downstream of the buffer volume and upstream of the FTIR cell, downstream of the FTIR cell and upstream of the UV–Vis cell, and downstream of the UV–Vis cell. Measurements were repeated three times under aerosol generation conditions. The aerodynamic and mobility diameters measured by the APS and SMPS were converted to geometric diameter using a particle density of 2.75 g $cm^{-3}$ and a dynamic shape factor of $\chi$=1.19. The resulting size distributions were expressed as particle numbers per size bin, normalized to the total number concentration, and averaged across the three repetitions. Relative decreases in individual size bins between sampling positions were used to qualitatively estimate particle losses. As measurements were conducted sequentially, identical aerosol generation conditions and particle concentrations could not be maintained across all positions. The experimental estimates were therefore compared with predictions from the Particle Loss Calculator software (PLC; Von der Weiden et al., 2009) to evaluate their consistency in terms of size dependence.

**Figure S.2a** compares the experimentally estimated losses with the PLC calculations. Despite variations in aerosol generation conditions between different sampling positions, the experimental measurements reproduce the general size dependence predicted by PLC, including enhanced losses for ultrafine and coarse particles**.** At the largest particle sizes, the experimentally estimated losses reached approximately 90%, whereas the PLC calculations approached nearly complete particle removal ∼100%. **Figure S.2b** presents the calculated cumulative losses along the different aerosol transport paths, distinguishing contributions upstream and downstream of the FTIR cell. Since the SMPS and APS were located downstream of the FTIR cell, only the losses occurring between the FTIR cell and the SMPS–APS sampling point were corrected.

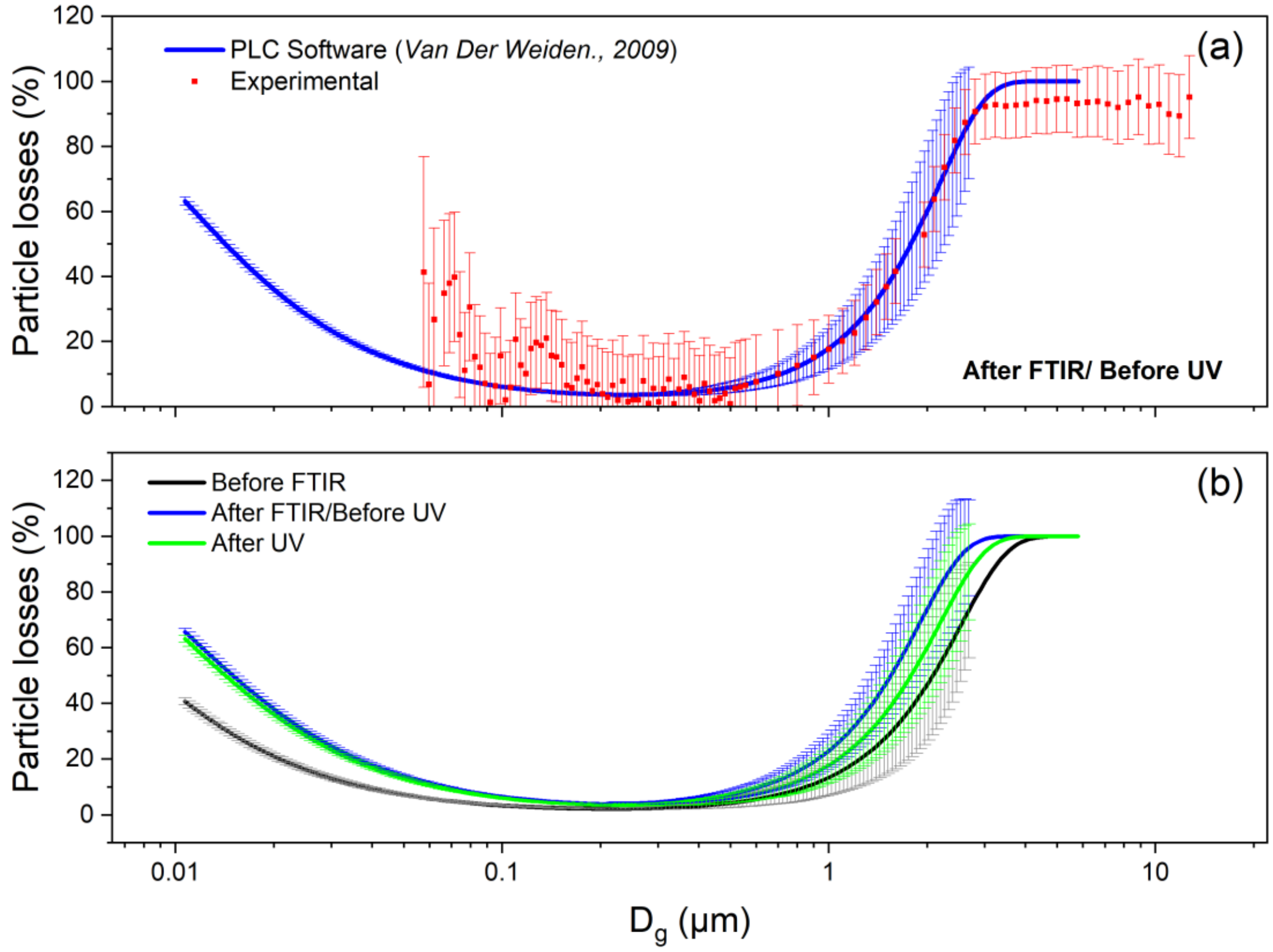

**Figure S.2:** Particle losses within the experimental aerosol setup as a function of geometric particle diameter $D_g$. **(a)** Comparison between particle losses estimated experimentally by positioning the SMPS-APS sampling system at different locations in the setup and losses calculated using the PLC software (Von der Weiden et al., 2009). Because the experimental measurements could not be performed under strictly identical aerosol-generation conditions, this comparison is considered qualitative. **(b)** Cumulative particle losses calculated using PLC for the aerosol transport paths in the experimental setup. Calculations account for diffusional, gravitational, and inertial deposition within the tubing system using the PLC software.

# Supplementary D: Effective density calculations

To compute the effective mixture density, a single representative density is assigned to each mineral as the average of the literature values listed in **Table S.3** (source https://webmineral.com/, last accessed Sept 2026).

**Table S.3:** Density ranges of major pure minerals composing mineral dust.

| Mineral group | Mineral | Density (g cm⁻³) |
|---|---|---|
| Phyllosilicates | Illite | 2.75 |
| Tectosilicates | Quartz | 2.65 |
| Tectosilicates | Feldspars | 2.59 |
| Carbonates | Calcite | 2.71 |
| Iron oxides | Hematite | 5.15 |

The mixture density $\rho_{mix}$ is calculated by the linear combination of mass fraction $w_i$ and density $\rho_i$ of the material $i$ with, and it is given by:

$$\rho_{mix} = \sum w_i \cdot \rho_i \qquad \text{(S.1)}$$

Mass fraction weights are used from Di Biagio et al., (2017). Assuming the uncertainties in the mineral mass fractions ($w_i$) and in the component densities ($\rho_i$) are independent, the combined propagated uncertainty in the mixture density ($\sigma_{\rho_{eff}}$) is 1%.  **Table S.4.** gives the mixing density and relative uncertainty for each dust sample.

**Table S.4:** Effective density and relative uncertainty used for each dust sample.

| Sample | Mixture density (g cm⁻³) | Relative uncertainty |
|---|---|---|
| ARIZONA | 2.75 | 1 % |
| ATACAMA | 2.76 | 1 % |
| CHINA | 2.72 | 1 % |
| NIGER | 2.76 | 1 % |
| SAUDI ARABIA | 2.70 | 1 % |
| TUNISIA | 2.74 | 1 % |

# Supplementary E: Sensitivity analysis on aerosols size modes

## E.1. Jacobian sensitivities

For the size distribution adjustment step between 1600 and 40000 cm-1, we calculate the Jacobian column sensitivities as shown in **fig S.4**, mathematically $\|\mathbf{K}_j\|_2$ , where **K** is the Jacobian matrix written by $\mathbf{K} = \Delta F/\Delta \mathrm{lnx}$ , and we test how strongly the modeled extinction changes when each retrieved parameter is perturbed by 1%.

The Jacobian column norms indicate that the first retrieved mode dominates the extinction sensitivity, particularly through its geometric standard deviation and geometric mean diameter. The second mode remains measurably sensitive, whereas the third coarse mode produces only weak direct spectral variations, indicating limited independent sensitivity over the selected inversion window. The Jacobian column norms indicate that the first retrieved mode is the most spectrally sensitive, particularly for the geometric standard deviation and geometric mean diameter of the fine mode, while sensitivity decreases progressively for coarse modes 1 and 2.

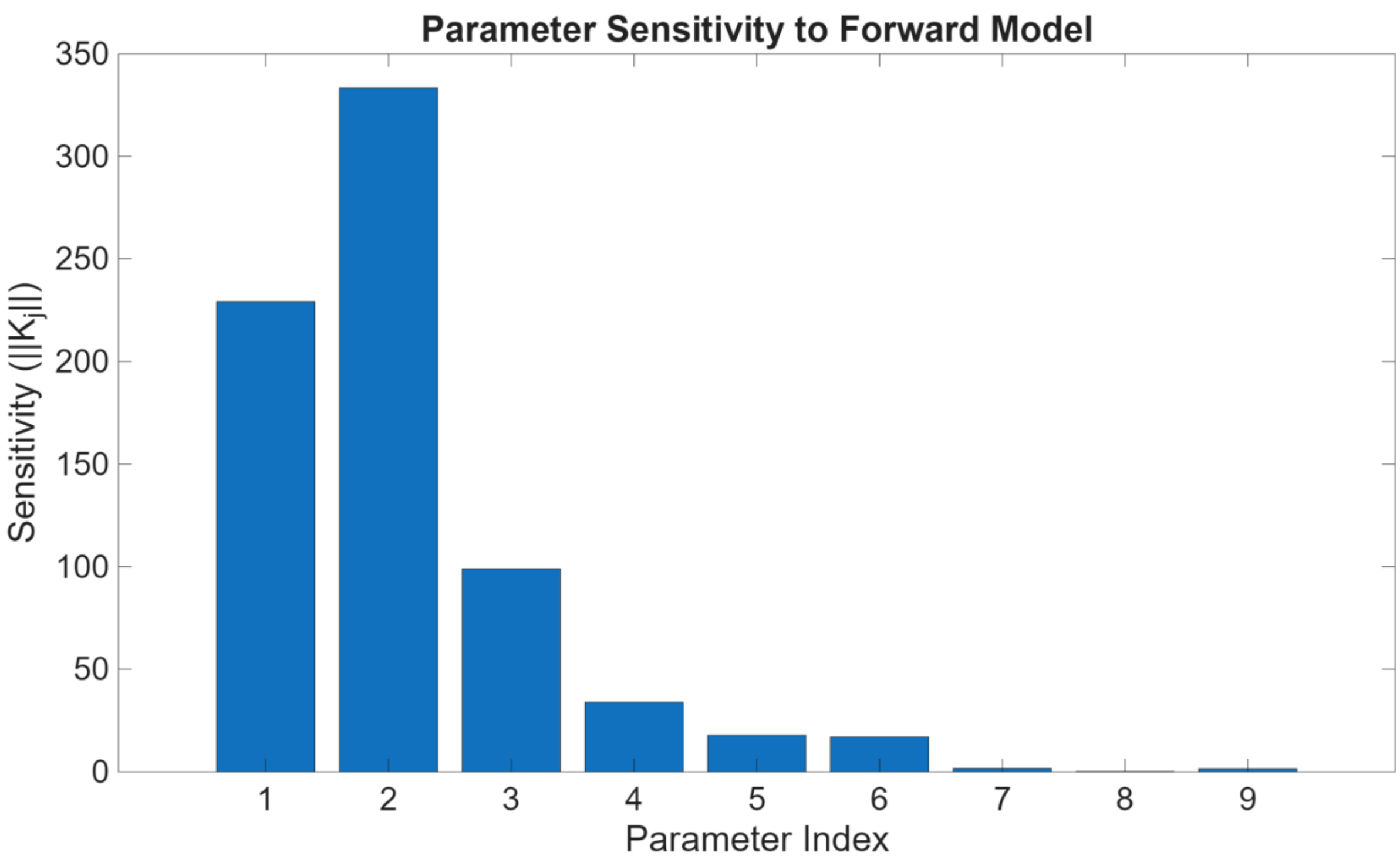

**Figure S.4:** Parameter sensitivity of the forward model. Indices 1–3 represent fine-mode parameters, indices 4–6 coarse mode 1, and indices 7–9 coarse mode 2, each ordered as the geometric mean diameter $\boldsymbol{D_g}$, geometric standard deviation $\boldsymbol{\sigma_g}$, and number concentration $\boldsymbol{N}$ respectively.

## E.2. Information content

The averaging Kernel matrix **A**, gives the sensitivity of the retrieved state to the true state, and is given by:

$$\mathbf{A} = \delta\hat{x}/\delta x = \mathbf{GK} \tag{S.2}$$

where **G** is the gain matrix which rows are the derivatives of the retrieved state with respect to the spectral points, and is written by:

$$\mathbf{G} = \delta\hat{x}/\delta \mathrm{y} = (\mathbf{K^T S_\varepsilon^{-1} K} + \mathbf{S_a^{-1}} + \boldsymbol{\gamma I})^{-1}\mathbf{K^T S_\varepsilon^{-1}} \tag{S.3}$$

Rodgers, 2000 showed that the trace of **A** represents the total degree of freedom for signal (DOFS), that gives the number of independent pieces of information provided by the observing system as regards the state vector. The retrieval information content was further quantified using the averaging kernel matrix as shown in **fig. S.5**. The diagonal elements of the averaging kernel are close to unity for most retrieved parameters, indicating that the inversion is predominantly driven by the measurements rather than by the *a priori* constraints. The total degree of freedom for signal (DOFS) is 8.07 for 9 retrieved parameters, indicating that the measurements provide approximately 8 independent pieces of information for the 9-parameter state vector, which demonstrates strong overall measurement information content. Only the geometric mean diameter of the second retrieved mode (CM1) exhibits reduced independent information content, with an averaging-kernel diagonal value significantly lower than unity. This indicates that although this parameter affects the extinction spectrum, its spectral signature is not fully independent from other retrieved quantities. This behavior is consistent with parameter coupling within the same lognormal mode. In particular, changes in geometric mean diameter and number concentration can generate partially similar extinction responses, leading to partial correlation between these parameters.

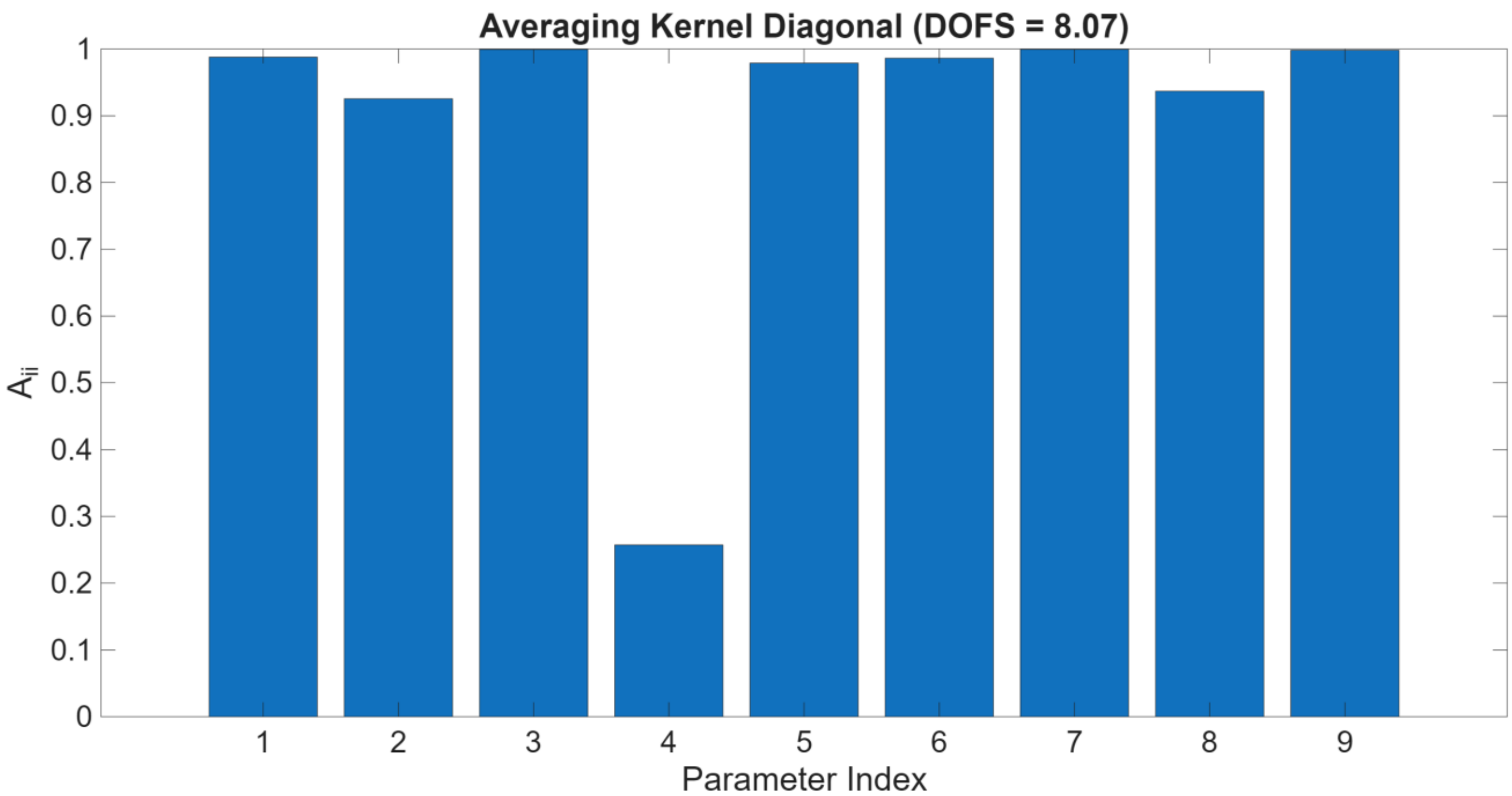

**Figure S.5:** Diagonal elements of the averaging kernel matrix for the retrieved size-distribution parameters. Indices 1–3 correspond to the fine mode ($\boldsymbol{D_g}$, $\boldsymbol{\sigma_g}$, $\boldsymbol{N}$), indices 4–6 to coarse mode 1, and indices 7–9 to coarse mode 2, with identical parameter order. Most parameters show averaging kernel values close to unity, indicating strong observational constraint, while coarse mode 1 geometric diameter (parameter 4) remains weakly constrained. The total degrees of freedom for signal (DOFS) is 8.07.

## E.3. Inverse Retrieval Methodology and Uncertainty Propagation

The optimal estimation method is used first in size distribution parameter adjustment and later in the spectral CRI retrieval. Below is the complete mathematical formulation governing both retrievals, parameter vectors, error covariance matrices, and total uncertainty propagation, summarized in **Table S.5**.

### E.3.1- A priori covariance matrix

For the size distribution adjustment step, we use the a priori covariance matrix $\mathbf{S}_{\boldsymbol{a}}$ that represents the uncertainty associated with the prior knowledge of the state vector before including the measurement information through the inversion process. It quantifies how strongly each retrieved parameter is allowed to vary around its prior value and therefore controls the regularization strength of the retrieval.

The a priori error covariance matrix $\mathbf{S}_{\boldsymbol{a}}$ is assumed diagonal with the $i^{\text{th}}$ diagonal element $\mathbf{S}_{\boldsymbol{a},\boldsymbol{ii}}$ defined as:

$$\mathbf{S}_{\boldsymbol{a},\boldsymbol{ii}} = \boldsymbol{\sigma}_{\boldsymbol{a},\boldsymbol{i}}^{\mathbf{2}} \text{ with } \sigma_{a,i} = x_{a,i}.\frac{p_{error,i}}{100} \quad \text{(S.4)}$$

where $p_{error,i}$ is the corresponding uncertainty on the prior knowledge of the aerosol size distribution parameters as estimated from the multimodal fit (between 20 and 30%) for the $i^{\text{th}}$ parameter of the state vector.

### E.3.2- Regularization covariance matrix

For the CRI retrieval step, the reference state vector is denoted as the regularization vector $\mathbf{x}_{\text{reg}}$. In accordance with the definitions of Rodgers (2000), $\mathbf{x}_{\text{reg}}$ is formulated as a regularization constraint rather than a strictly measurement-independent Bayesian a priori. $\mathbf{x}_{\text{reg}}$ is obtained iteratively starting from Rayleigh scattering simulation than iterative convergence with respect to the spectral measurement.

The regularization constraint matrix $\mathbf{S}_{\text{reg}}$ assumes a constant relative prior uncertainty $\boldsymbol{p}_{reg}(\nu)$ = 5% applied uniformly across the entire spectral range. The diagonal variance elements are given by

$$\mathbf{S}_{\boldsymbol{reg},\boldsymbol{ii}} = \sigma_{reg,i}^{2}(\nu), \text{ with } \sigma_{\text{reg},i}(\nu) = x_{reg,i}.\frac{p_{reg,i}(\nu)}{100} \quad \text{(S.5)}$$

### E.3.3- Measurement error covariance matrix

The measurement error covariance matrix is influenced by the radiometric calibration and the radiometric noise, given by the signal-to-noise ratio (SNR). This error covariance matrix is also assumed to be diagonal, and the $i^{\text{th}}$ diagonal element can be computed as follows:

$$\mathbf{S}_{\boldsymbol{e},\boldsymbol{ii}} = \sigma_{smooth,i}^{2}(\nu), \text{ with } \sigma_{\text{smooth},i}(\nu) = \left|\beta_{\text{smoothed},i}(\nu) - \beta_{\text{raw},i}(\nu)\right| \quad \text{(S.6)}$$

### E.3.4- Non-retrieved parameters error covariance matrix $S_b$

The covariance matrix $\mathbf{S}_{\mathbf{b}}$ accounts for uncertainties associated with the fixed parameters of the CRI retrieval.

In the case of the size distribution adjustment, the uncertainties associated with the fixed vector **b** are neglected because the ultrafine mode has negligible spectral influence over the inversion interval and therefore does not significantly contribute to the total error.

In the case of the CRI retrieval, it corresponds to the multimodal aerosol size-distribution parameters. For all modes, the parameters $\left(D_{g,i}, \sigma_{g,i}, N_i\right)$ of the $i^{th}$ mode are treated as fixed elements of the non-retrieved parameter vector **b**. The covariance matrix $\mathbf{S}_{\mathbf{b}}$ is directly taken as the posterior covariance matrix $\mathbf{S}_{\mathbf{x},\mathbf{SD}}$ obtained from the size-distribution adjustment, thereby propagating the uncertainty of the adjusted multimodal aerosol parameters into the CRI retrieval.

### E.3.5- The total error covariance matrix

The posterior knowledge of the retrieved state vector is described by the total error covariance matrix $\mathbf{S_x}$, and can be written as:

$$\begin{cases} \mathbf{S}_{\mathbf{x,SD}} = \mathbf{S}_{\text{smoothing}} + \mathbf{S}_{\text{meas.}} \\ \mathbf{S}_{\mathbf{x,CRI}} = \mathbf{S}_{\text{reg. contr}} + \mathbf{S}_{\text{meas.}} + \mathbf{S}_{\text{fwd.mod.}} \end{cases} \quad (S.7)$$

where $\mathbf{S}_{\text{smoothing}}$ represents the uncertainty introduced by the a priori constraint and quantifies the residual influence of the prior information on the retrieved CRI. It is given by:

$$\mathbf{S}_{\text{smoothing}} = (\mathbf{A} - \mathbf{I})\mathbf{S}_{\text{a}}(\mathbf{A} - \mathbf{I})^{\text{T}} \quad (S.8)$$

$\mathbf{S}_{\text{meas.}}$ is the contribution of the measurement error covariance $\mathbf{S}_{\text{m}}$ associated with spectral noise, and it is written as:

$$\mathbf{S}_{\text{meas.}} = \mathbf{G}\mathbf{S}_{\text{e}}\mathbf{G}^{\mathbf{T}} \quad (S.9)$$

$\mathbf{S}_{\text{smoothing}}$ represents the uncertainty introduced by the regularization constraint and quantifies the residual influence of the regularization information. It is given by:

$$\mathbf{S}_{\text{reg. contr}} = (\mathbf{A} - \mathbf{I})\mathbf{S}_{\text{reg}}(\mathbf{A} - \mathbf{I})^{\text{T}} \quad (S.10)$$

$\mathbf{S}_{\text{fwd.mod.}}$ is the contribution of the forward model error covariance matrix $\mathbf{S}_{\text{f}}$ associated with uncertainties from non-retrieved model parameters described by the covariance matrix $\mathbf{S}_{\text{b}}$**:**

$$\mathbf{S}_{\text{fwd.mod.}} = \mathbf{G}\mathbf{K}_{\text{b}}\mathbf{S}_{\text{b}}(\mathbf{G}\mathbf{K}_{\text{b}})^{\mathbf{T}} = \mathbf{G}\mathbf{S}_{\text{f}}\mathbf{G}^{\mathbf{T}} \quad (S.11)$$

where $\mathbf{K}_{\text{b}}$ is the Jacobian of the forward model with respect to the non-retrieved parameter vector **b**. Note that **A** is the averaging Kernel matrix and **G** is the gain matrix defined in the **equation (S.2)** and **(S.3),** respectively.

**Table S.5:** Parameters and equations summary of the retrieval process.

| Inversion | Symbol | Size distribution adjustment | Inversion | Symbol | CRI retrieval |
|---|---|---|---|---|---|
| **State vector** | **x** | $\mathbf{x}_{\mathbf{SD}} = [D_{g,i}, \sigma_{g,i}, N_i, ..]^{\mathbf{T}}$, *i*= fine, coarse- 1 and coarse-2 | **State vector** | **x** | $\mathbf{x}_{\mathbf{reg}} = [n(\nu), \kappa(\nu)]^{\text{T}}$ |
| **A priori** | $\mathbf{x}_{\boldsymbol{a}}$ | Initial size distribution modes parameters | **Regularization vector** | $\mathbf{x}_{\mathbf{reg}}$ | Iteratively converged $[n(\nu), \kappa(\nu)]^{\text{T}}$ |
| **A priori covariance matrix** | $\mathbf{S}_{\mathbf{a}}$ | $\mathbf{S}_{\boldsymbol{a},\boldsymbol{ii}} = \sigma_{a,i}^2$ , *assumed diagonal* with $\sigma_{a,i} = x_{a,i} \cdot \frac{p_{error,i}}{100}$ | **Regularization covariance matrix** | $\mathbf{S}_{\mathbf{reg}}$ | $\mathbf{S}_{\boldsymbol{reg},\boldsymbol{ii}} = \sigma_{reg,i}^2(\nu)$, *assumeddiagonal*with $\sigma_{reg,i}(\nu) = x_{reg,i} \cdot \frac{p_{reg,i}(\nu)}{100}$ |
| **Relative a priori uncertainty** | $\boldsymbol{p}_{\boldsymbol{error},\boldsymbol{i}}$ | 20 *to* 30 % | **Relative Regularization uncertainty** | $\boldsymbol{p}_{\boldsymbol{reg},\boldsymbol{i}}(\nu)$ | 5 % |
| **Measurement vector** | **Y** | Measured extinction spectrum | | | |
| **Measurement covariance matrix** | $\mathbf{S}_{\mathbf{e}}$ | $\mathbf{S}_{\mathbf{e},\boldsymbol{ii}} = \boldsymbol{\sigma}_{\boldsymbol{smooth},\boldsymbol{i}}^{\mathbf{2}}(\nu)$ , assumed diagonal with $\boldsymbol{\sigma}_{\mathbf{smooth},\boldsymbol{i}}(\nu) = \lvert \beta_{\text{smoothed},i}(\nu) - \beta_{\text{raw},i}(\nu) \rvert$ | | | |
| **Non-retrieved parameters** | **b** | $\mathbf{b}_{\boldsymbol{UFM}} = [D_{g,UFM}, \sigma_{g,UFM}, N_{UFM}]^{\mathbf{T}}$, *i*= ultra-fine mode | | | $\mathbf{b} = [D_{g,i}, \sigma_{g,i}, N_i, ...]^{\mathbf{T}}$ , *i*= all modes |
| **Non-retrieved parameters covariance matrix** | $\mathbf{S}_{\mathbf{b}}$ | Not propagated. Negligible ultrafine-mode sensitivity over the spectral range used for the adjustment. | | | $\mathbf{S}_{\mathbf{b}} = \mathbf{S}_{\mathbf{tot,SD}}$ |

| | | | |
|---|---|---|---|
| **Retrieved state covariance** | $\mathbf{S_x}$ | $\mathbf{S_{x,SD}} = \mathbf{S_{smoothing}} + \mathbf{S_{meas}}$ | $\mathbf{S_{x,CRI}} = \mathbf{S_{reg.contr}} + \mathbf{S_{meas}} + \mathbf{S_{fwd,mod}}$ |
| **Reported uncertainty** | $\sigma_{\mathbf{x}}$ | $\sigma_{x,i,SD} = \sqrt{\mathbf{S_{x,SD,ii}}}$ | $\sigma_{x,CRI} = \sqrt{\mathbf{S}_{\mathbf{x,CRI},ii}}$ |

# Supplementary F: Spectral measurements errors

## F.1. Spectral noise

**Table S.6:** Instrumental noise in FIR, MIR and UV-VIS spectrometers.

| Spectral region | Spectral range ($cm^{-1}$) | Instrumental noise ($m^{-1}$) |
|---|---|---|
| FIR | 100-650 | $5 \times 10^{-5}$ |
| MIR | 650-4000 | $5 \times 10^{-5}$ |
| UV–Vis | 10000-40000 | $2.3 \times 10^{-4}$ |

## F.2. Smoothing errors

The extinction spectrum uncertainty was estimated by combining instrumental and processing contributions on the native spectral grid of the smoothed extinction spectrum (1 $cm^{-1}$) was linearly interpolated in order to avoid artificial oversampling of the processed data. The applied smoothing correction was quantified as the spectral difference between smoothed and raw extinction coefficients, ($\Delta\beta(\nu) = \beta_{smoothed}(\nu) - \beta_{raw}(\nu)$). Instrumental uncertainty was assigned independently to each measured spectral domain using representative absolute uncertainties for FTIR measurements in the FIR–MIR range and UV-Vis measurements in the UV-Vis range. Smoothing-induced uncertainty was estimated as 1 % of the absolute applied correction. In the unmeasured SWIR interval (4000–10000 $cm^{-}$1) no direct measurement uncertainty was assigned, and this interval was excluded from direct uncertainty interpretation. Total uncertainty was obtained by quadratic propagation of instrumental and processing contributions. The uncertainty analysis is summarized in four panels of **Fig. S.6** :(a) Raw and smoothed extinction spectra shown only in measured spectral intervals. (b) Applied smoothing correction $\Delta\beta(\nu)$. (c) Instrumental $\sigma_{instr}$, smoothing-induced $\sigma_{smooth}$, and total uncertainty contributions $\sigma_{tot}$. (d) Relative total uncertainty expressed as percentage of extinction.

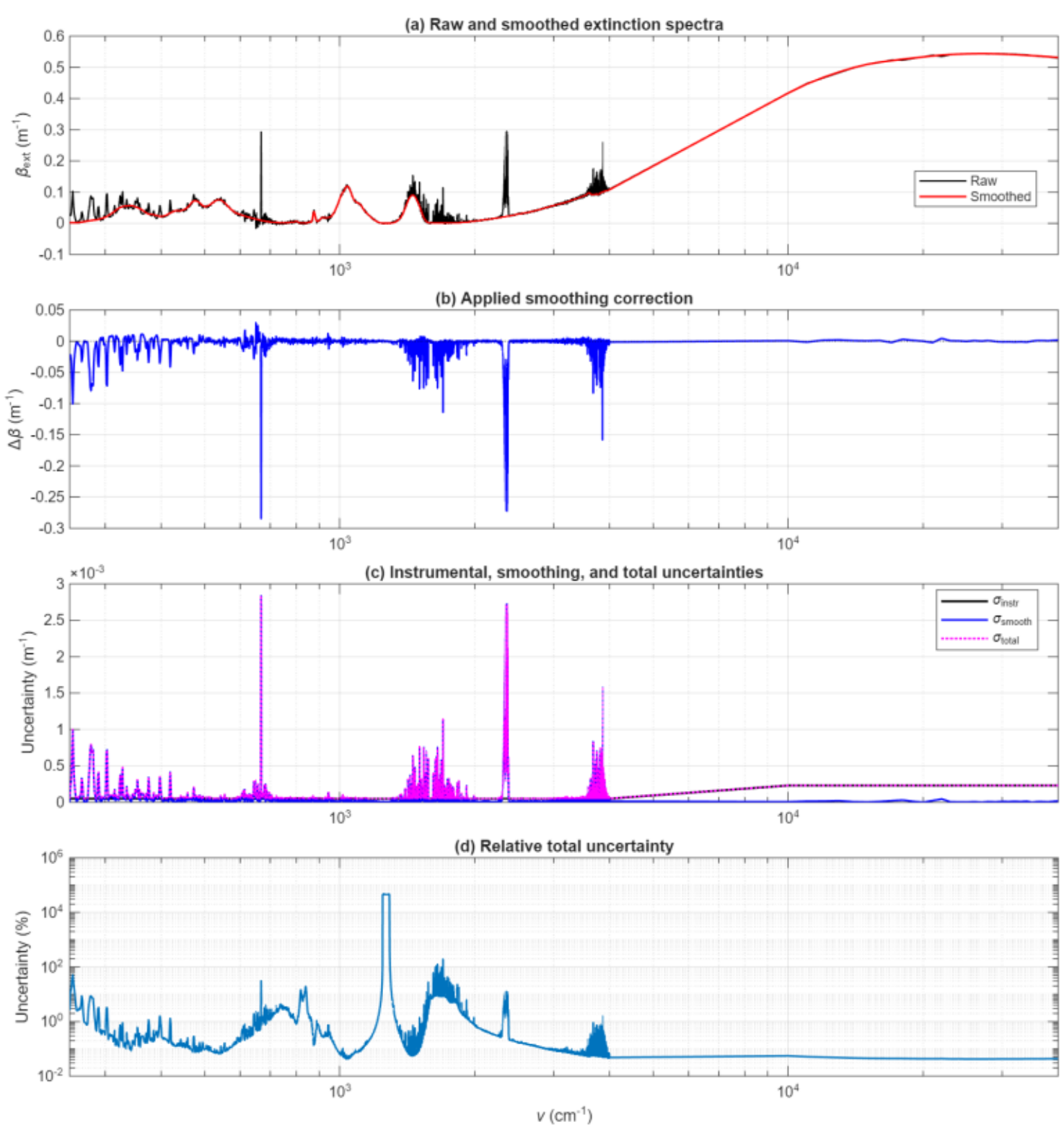

**Figure S.6: S**pectral pre-processing and associated uncertainty analysis for the Tunisia extinction spectrum. (a) Raw measured extinction spectrum (black) compared with the smoothed spectrum (red). (b) Applied smoothing correction $\boldsymbol{\Delta\beta(\nu)}$ , defined as the difference between raw and smoothed spectra. (c) Instrumental uncertainty $\boldsymbol{\sigma_{instr}}$ , smoothing uncertainty $\boldsymbol{\sigma_{smooth}}$, and resulting total uncertainty $\boldsymbol{\sigma_{total}}$ . (d) Relative total uncertainty (%) across the investigated spectral range.

The uncertainty analysis shows that the extinction spectrum remains globally well constrained over the measured FIR, MIR, and UV-Vis regions, with median relative uncertainties of 0.19 %, 0.16 %, and 0.04 %, respectively. Instrumental uncertainty dominates most of the spectrum, whereas smoothing-induced uncertainty becomes significant only near gases perturbations spectral features, particularly water vapor, and carbon dioxide. Although local relative uncertainty spikes are observed, these occur where extinction approaches zero near 1300 cm$^{-1}$ and therefore do not affect the overall uncertainty of the spectrum. The UV-Vis region exhibits the highest stability, while the SWIR interval is not measured and is therefore excluded from direct interpretation.

# Supplementary G: Size distribution measurements errors

**Table S.7:** Relative uncertainties sources of the SMPS and APS instruments.

| Instrument | Uncertainty source | Relative uncertainty | Reference |
|---|---|---|---|
| SMPS | Raw counts $c_i$ | $100/\sqrt{c_i}$ | Buonanno et al., 2009 |
| | CPC aerosol-flow | 5% | TSI Incorporated, 2007 |
| | CPC counting efficiency | 8% | Buonanno et al., 2009; Hermann et al., 2007 |
| | Diffusion-correction efficiency | 10% | Buonanno et al., 2009; Hermann et al., 2007 |
| | Geometric Diameter $D_g$ | 12% | Propagated |
| APS | Raw counts $c_i$ | $100/\sqrt{c_i}$ | Buonanno et al., 2009 |
| | APS aerosol-flow | 10% | TSI Incorporated, 2012 |
| | Geometric Diameter $D_g$ | 6% | Propagated |

# Supplementary H: Pure minerals molecular assignments list

**Table S.8:** Molecular assignment for main minerals composing aerosols found in literature.

| Mineral | Wavenumber (cm$^{-1}$) | Molecular Assignment | References |
|---|---|---|---|
| **Kaolinite Muscovite, Illite** | 1037, 1033 | Si-O-Si asymmetrical stretching | Balan et al., 2001; Madejová et al., 2017 |
| | 1116 to 1121 | Si-O symmetrical stretching | |
| | 910 | -O-H bending | Farmer, 1974; Madejová et al., 2017 |
| | 755 to 697 | Al-O-Si bending | |
| | 540 | Al-O-Si bending | |
| | 472 | Si-O-Si bending | |
| | 428 | Si-O-Si bending | |
| | 347, 278 | | Glotch et al., 2007 |
| | 506 and 788 | Al-O-Si bending | Balan et al., 2001; Chehab et al., 2024 |
| | 253 | -O-H liberation | |
| **Calcite** | 1430 | -C-O asymmetrical stretching | Campbell and Poduska, 2020 |
| | 875 | -C-O bending | |
| | 712 | -C-O bending | |
| | 300 | Ca-O lattice mode | |
| **Quartz** | 1080, 1102, 1177 | Si-O asymmetrical stretching | Müller et al., 2014 |
| | 780, 800 | Si-O symmetrical stretching | |
| | 695 | Si-O-Si symmetrical bending | |
| | 520 | Si-O-Si asymmetrical bending | |
| | 450 | Si-O-Si bending | |
| **Hematite** | 560 | Fe-O stretching | Glotch et al., 2006 |
| | 456 | Fe-O bending | |
| | 390 | Fe-O stretching | |
| | 319 | Fe-O stretching | |

# Supplementary I: Size distribution adjustment results

**Figure S.6** compares the measured extinction spectra with the spectra fitted after optimization of the aerosol size-distribution parameters using the OEM procedure described in **Section 2.5.5**. The retrieved size distribution parameters are found in **Table S.8**. For all samples except Niger, the inversion is performed from 1600 to 40000 cm$^{-1}$ in order to exclude the carbonate absorption band centered near 1400 cm$^{-1}$**,** since this feature originates from mineral

absorption and cannot be reproduced by particle-size adjustment alone. For the Niger sample, the fitting interval is extended down to 1300 cm⁻¹ because no detectable carbonate absorption is observed in this spectral region.

Overall, the optimized size distributions reproduce the measured extinction spectra consistently over the UV–Vis–MIR range for all dust samples, indicating that the selected lognormal-mode adjustments capture the dominant spectral variability associated with particle size. The agreement between measured and fitted spectra is estimated by using the relative difference between the two spectra. The relative difference values range from 0.43 % to 1.83 %, showing that residual differences remain below 2 % for all samples despite the spectral contrast between source regions. The best agreement is obtained for the China sample (0.43 %**)** and Arizona (0.66 %**),** whereas slightly larger residuals are observed for Saudi Arabia (1.83 %) and Atacama (1.44 %). Although, some spectral structures remain unresolved near 3800 cm⁻¹ and are visible in all samples, with stronger differences for Atacama, Niger, and Saudi Arabia. These residual features are attributed to OH- stretching vibrations and are not reproduced by the fit because the inversion modifies only the lognormal size distribution parameters while keeping the complex refractive index fixed. In addition, the refractive-index dataset from Di Biagio et al., (2019) does not include measured optical constants around 3800 cm⁻¹, where extrapolated values are used instead. This limitation directly affects the representation of absorption in this spectral region and explains the remaining mismatch between measured and fitted spectra.

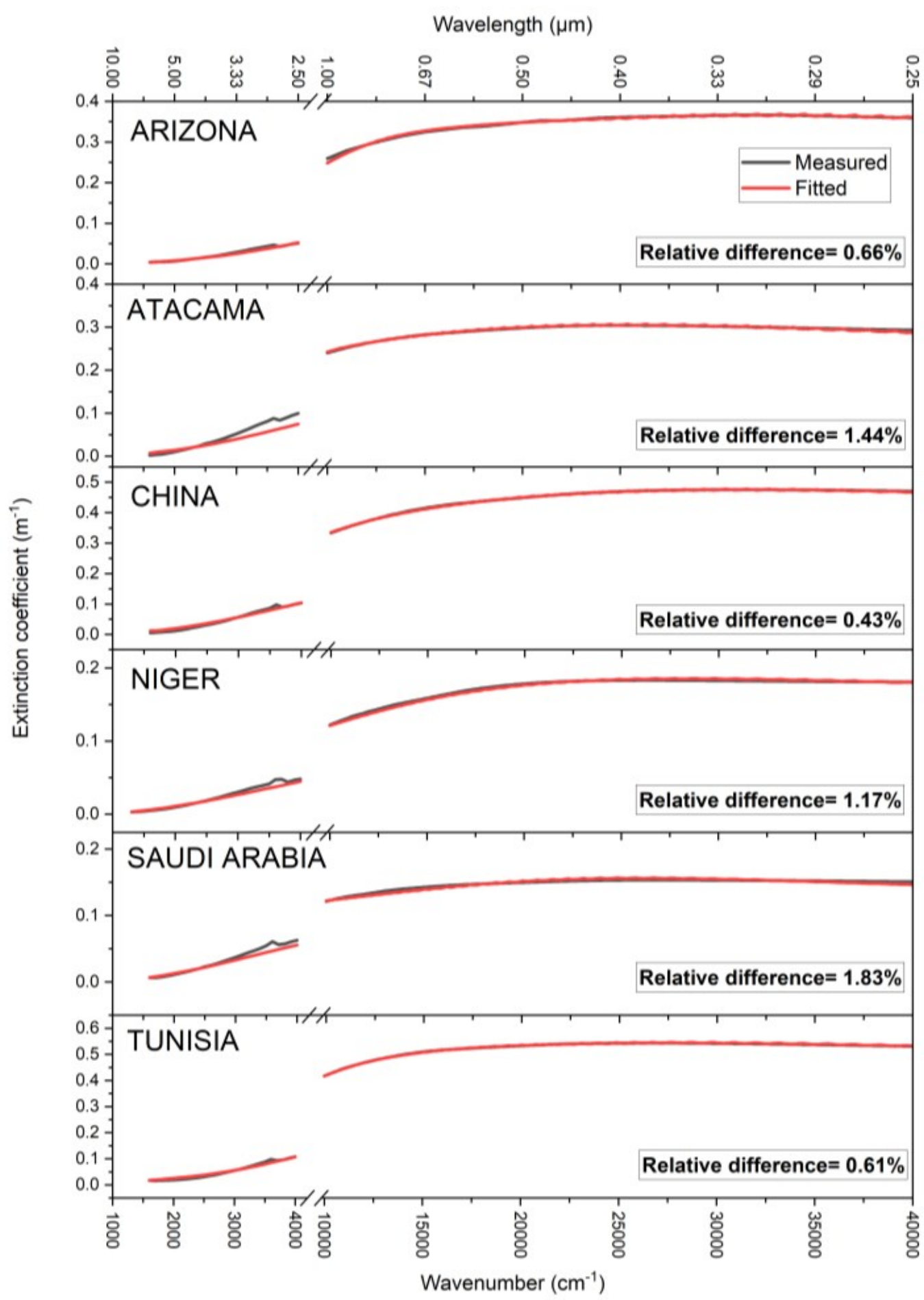

**Figure S.6:** Measured (black) and fitted (red) extinction spectra after aerosol size distribution adjustment using the OEM inversion for six mineral dust samples Arizona, Atacama, China, Niger, Saudi Arabia and Tunisia. The inversion is performed over 1600–40000 cm⁻¹, except for Niger (1300–40000 cm⁻¹) due to the absence of detectable carbonate absorption near 1400 cm⁻¹.

**Table S.9:** Parameters of the multimodal particle size distribution fits and their adjustment the six investigated dust samples. For each sample and particle-size mode, the table reports the fitted parameter obtained from the multimodal size-distribution fit, the corresponding value after retrieval adjustment, the associated retrieval uncertainty, and the relative shift to the initial fitted value.

| | | | **ARIZONA** | | | |
|---|---|---|---|---|---|---|
| | **Mode** | **Parameter** | **Multimodal fit** | **Adjustment retrieval** | **Retrieval error** | **Relative shift** |
| Retrieved | FM | *Dg (µm)* | 0.273 | 0.223 | 2% | 18% |
| | | *σ* | 1.760 | 1.913 | 1% | 9% |
| | | *N (#/cm³)* | 128080 | 272058 | 2% | 112% |
| | CM1 | *Dg (µm)* | 0.810 | 0.920 | 4% | 14% |
| | | *σ* | 1.550 | 1.350 | 1% | 13% |
| | | *N (#/cm³)* | 7295 | 7490 | 19% | 3% |
| | CM2 | *Dg (µm)* | 6 | 3.24 | 9% | 46% |
| | | *σ* | 1.45 | 1.3 | 2% | 10% |
| | | *N (#/cm³)* | 18 | 20 | 12% | 11% |
| Fixed | UFM | *Dg (µm)* | 0.013 | | | |
| | | *σ* | 1.8 | | | |
| | | *N (#/cm³)* | 5714 | | | |
| | | | **ATACAMA** | | | |
| | **Mode** | **Parameter** | **Multimodal fit** | **Adjustment retrieval** | **Retrieval error** | **Relative shift** |
| *Retrieved* | *FM* | *Dg (µm)* | 0.350 | 0.29127 | 2% | 17% |
| | | *σ* | 1.600 | 1.5906 | 2% | 1% |
| | | *N (#/cm³)* | 71995 | 116180 | 5% | 61% |
| | *CM1* | *Dg (µm)* | 0.690 | 0.81938 | 4% | 19% |
| | | *σ* | 1.620 | 1.6021 | 1% | 1% |
| | | *N (#/cm³)* | 11417 | 16128 | 13% | 41% |
| | *CM2* | *Dg (µm)* | 4 | 1.1348 | 13% | 72% |
| | | *σ* | 1.5 | 1.9633 | 4% | 31% |
| | | *N (#/cm³)* | 8 | 20 | 29% | 150% |
| *Fixed* | *UFM* | *Dg (µm)* | 0.02 | | | |
| | | *σ* | 1.6 | | | |
| | | *N (#/cm³)* | 561 | | | |
| | | | **CHINA** | | | |
| | **Mode** | **Parameter** | **Multimodal fit** | **Adjustment retrieval** | **Retrieval error** | **Relative shift** |
| *Retrieved* | *FM* | *Dg (µm)* | 0.340 | 0.24733 | 3% | 27% |
| | | *σ* | 1.640 | 1.6908 | 3% | 3% |
| | | *N (#/cm³)* | 141204 | 255920 | 7% | 81% |
| | *CM1* | *Dg (µm)* | 0.770 | 0.75632 | 5% | 2% |
| | | *σ* | 1.560 | 1.6204 | 1% | 4% |
| | | *N (#/cm³)* | 9778 | 26111 | 15% | 167% |
| | *CM2* | *Dg (µm)* | 4 | 2.0582 | 21% | 49% |
| | | *σ* | 1.4 | 1.4281 | 6% | 2% |
| | | *N (#/cm³)* | 28 | 22.388 | 67% | 20% |
| *Fixed* | *UFM* | *Dg (µm)* | 0.02 | | | |
| | | *σ* | 1.75 | | | |
| | | *N (#/cm³)* | 6558 | | | |

| NIGER | | | | | | |
|---|---|---|---|---|---|---|
| | **Mode** | **Parameter** | **Multimodal fit** | **Adjustment retrieval** | **Retrieval error** | **Relative shift** |
| *Retrieved* | *FM* | *Dg (µm)* | 0.37 | 0.2404569 | 2% | 35% |
| | | *σ* | 1.74 | 1.8877573 | 1% | 8% |
| | | *N (#/cm³)* | 55340 | 126189.98 | 2% | 128% |
| | *CM1* | *Dg (µm)* | 1.11 | 1.5131387 | 2% | 36% |
| | | *σ* | 1.6 | 1.3667767 | 1% | 15% |
| | | *N (#/cm³)* | 4714 | 1081.1383 | 7% | 77% |
| | *CM2* | *Dg (µm)* | 10 | 3.510181 | 1% | 65% |
| | | *σ* | 1.38 | 1.2972866 | 0% | 6% |
| | | *N (#/cm³)* | 1 | 20 | 1% | 1900% |
| *Fixed* | *UFM* | *Dg (µm)* | 0.02 | | | |
| | | *σ* | 1.62 | | | |
| | | *N (#/cm³)* | 8506 | | | |

| SAUDI ARABIA | | | | | | |
|---|---|---|---|---|---|---|
| | **Mode** | **Parameter** | **Multimodal fit** | **Adjustment retrieval** | **Retrieval error** | **Relative shift** |
| *Retrieved* | *FM* | *Dg (µm)* | 0.35 | 0.19698 | 9% | 44% |
| | | *σ* | 1.85 | 1.944 | 9% | 5% |
| | | *N (#/cm³)* | 47496 | 1.08E+05 | 16% | 128% |
| | *CM1* | *Dg (µm)* | 0.75 | 0.79481 | 10% | 6% |
| | | *σ* | 1.5 | 1.5629 | 3% | 4% |
| | | *N (#/cm³)* | 4665 | 6310.8 | 52% | 35% |
| | *CM2* | *Dg (µm)* | 5.5 | 6.2963 | 3% | 14% |
| | | *σ* | 1.3 | 1.2999 | 3% | 0% |
| | | *N (#/cm³)* | 11 | 20 | 7% | 82% |
| *Fixed* | *UFM* | *Dg (µm)* | 0.01 | | | |
| | | *σ* | 1.8 | | | |
| | | *N (#/cm³)* | 7294 | | | |

| TUNISIA | | | | | | |
|---|---|---|---|---|---|---|
| | **Mode** | **Parameter** | **Multimodal fit** | **Adjustment retrieval** | **Retrieval error** | **Relative shift** |
| *Retrieved* | *FM* | *Dg (µm)* | 0.350 | 0.21892 | 2% | 37.5% |
| | | *σ* | 1.650 | 1.96821 | 1% | 19.3% |
| | | *N (#/cm³)* | 191658 | 360535.11 | 3% | 88.1% |
| | *CM1* | *Dg (µm)* | 0.880 | 0.87353 | 2% | 0.7% |
| | | *σ* | 1.540 | 1.42308 | 1% | 7.6% |
| | | *N (#/cm³)* | 15324 | 15443.523 | 12% | 0.8% |
| | *CM2* | *Dg (µm)* | 7 | 7.08837 | 5% | 1.3% |
| | | *σ* | 1.5 | 1.49987 | 7% | 0.0% |
| | | *N (#/cm³)* | 4 | 20.75124 | 13% | 418.8% |
| *Fixed* | *UFM* | *Dg (µm)* | 0.02 | | | |
| | | *σ* | 1.77 | | | |
| | | *N (#/cm³)* | 1199 | | | |

# Supplementary J: CRI sensitivity to errors obtained from the measured and adjusted size distributions

**Figure S.6** shows Arizona CRI errors sensitivity on the size-distribution adjustment. The propagated CRI uncertainty is strongly wavelength dependent. The largest differences between the uncertainty envelopes obtained from the measured and adjusted size distributions occur in the infrared spectral range, whereas their influence becomes considerably smaller toward the UV–visible. This indicates that the retrieval is more sensitive to the treatment of the particle size distribution in the infrared part of the spectrum.

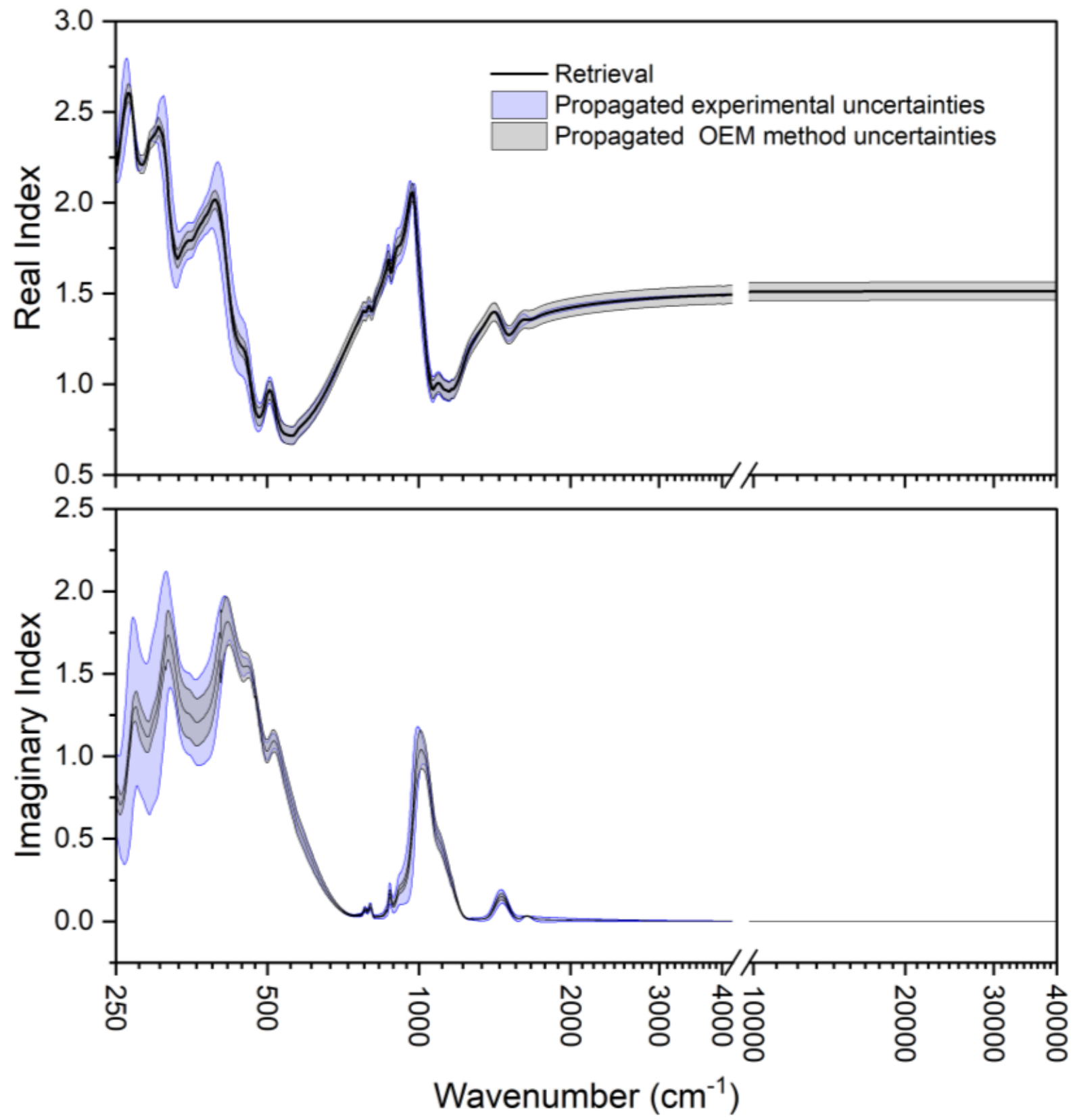

**Figure S.6:** Sensitivity of the retrieved complex refractive index uncertainty to the treatment of the particle size distribution, illustrated for the Arizona sample. Retrieved (upper pannel) real, and (lower pannel) imaginary refractive indices are shown together with their uncertainty envelopes obtained by propagating either the experimental uncertainties associated with the measured particle size distribution (blue envelope) and those associated with the OEM adjusted multimodal fit used in the retrieval (gray envelope).

# Supplementary K: Comparison with Di Biagio et al., 2017

## K.1- Size distributions comparison

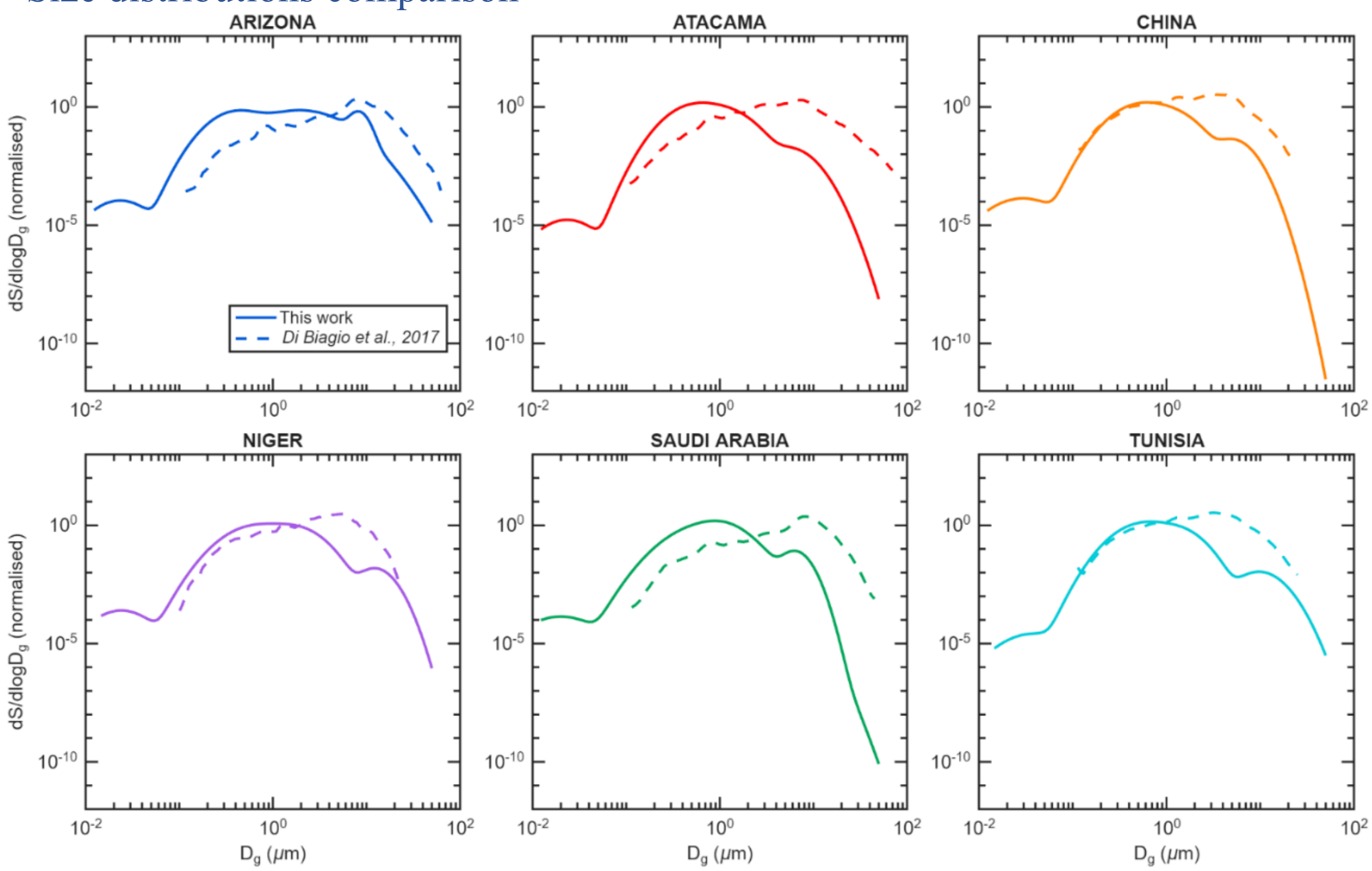

**Figure S.8:** Surface size distribution comparison between this study and Di Biagio et al., 2017**.**

## K.2- Extinction coefficient comparison

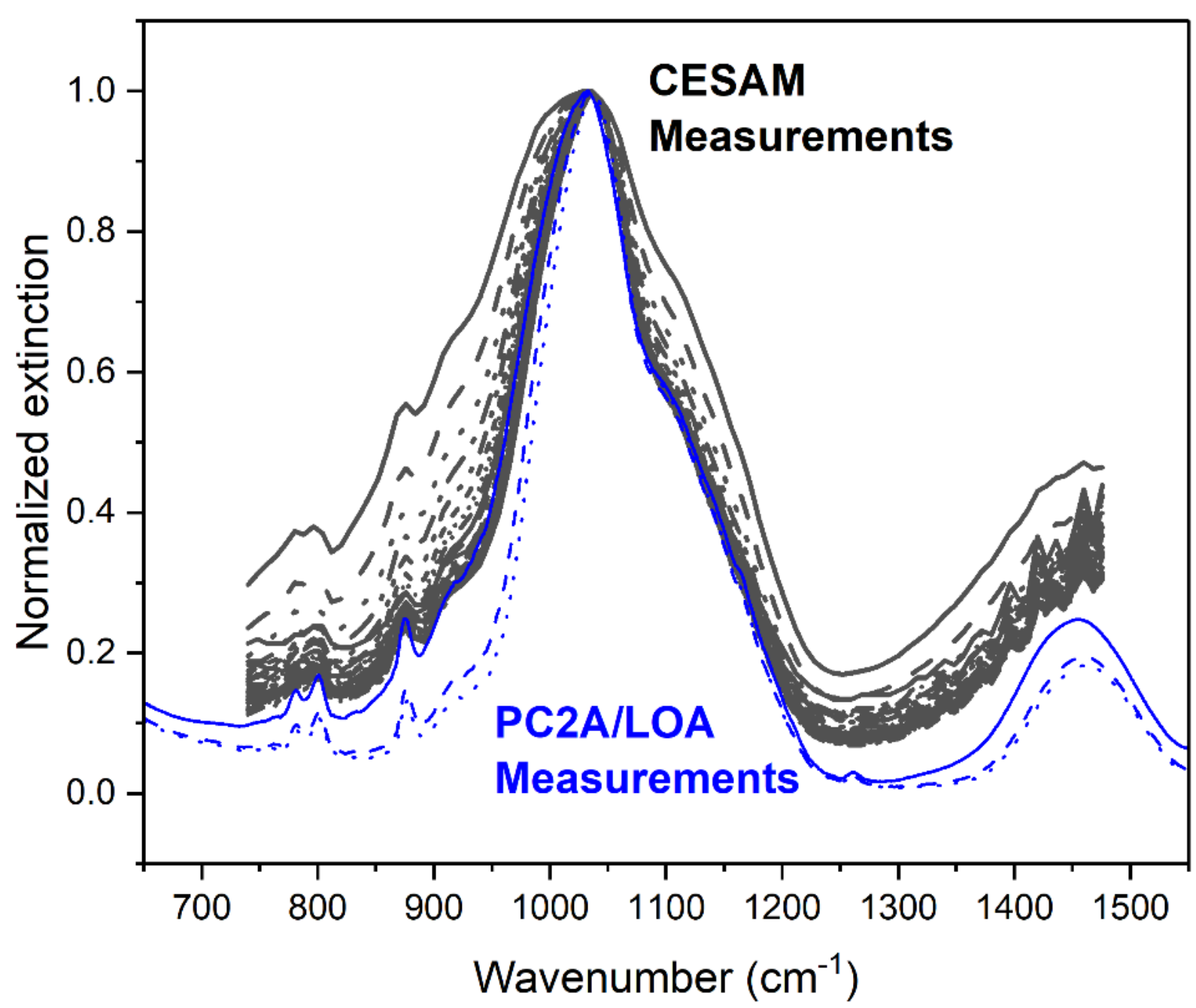

**Figure S.9:** Normalized extinction evolution comparison of Arizona sample between this study performed at PC2A/LOA platform and Di Biagio et al., 2017 at CESAM chamber.